**Nitrate ions spikes in ice cores are not suitable proxies
for solar proton events**

Katharine A. Duderstadt, Jack E. Dibb, Nathan A. Schwadron, and Harlan E. Spence,
Institute for the Study of Earth, Oceans, and Space, University of New Hampshire,
Durham, New Hampshire, USA.

Stanley C. Solomon, National Center for Atmospheric Research, Boulder, Colorado,
USA.

Valery A. Yudin, 1. National Center for Atmospheric Research, Boulder, Colorado, USA.
2. CIRES, Space Weather Prediction Center, University of Colorado, Boulder, Colorado,
USA.

Charles H. Jackman, NASA Goddard Space Flight Center, Greenbelt, Maryland, USA.

Cora E. Randall, Laboratory for Atmospheric and Space Physics, University of Colorado,
Boulder, Colorado, USA.  Department of Atmospheric and Oceanic Sciences, University
of Colorado Boulder, Boulder, Colorado, USA.

Corresponding author: K. A. Duderstadt, Earth Systems Research Center, University of
New Hampshire, Durham, NH 03824, USA. (duderstadtk@gust.sr.unh.edu)

**Key Points**:
- SPE-enhanced atmospheric $NO_y$ does not account for $NO_3^-$ spikes in ice cores.
- $NO_3^-$ in ice cores is a poor proxy for historical SPEs.

**Abstract**

Nitrate ion spikes in polar ice cores are contentiously used to estimate the intensity, frequency, and probability of historical solar proton events, quantities that are needed to prepare for potentially society-crippling space weather events. We use the Whole Atmosphere Community Climate Model to calculate how large an event would have to be to produce enough odd nitrogen throughout the atmosphere to be discernible as nitrate peaks at the Earth's surface. These hypothetically large events are compared with probability of occurrence estimates derived from measured events, sunspot records, and cosmogenic radionuclides archives. We conclude that the fluence and spectrum of solar proton events necessary to produce odd nitrogen enhancements equivalent to the spikes of nitrate ions in Greenland ice cores are unlikely to have occurred throughout the Holocene, confirming that nitrate ions in ice cores are not suitable proxies for historical individual solar proton events.

# 1. Introduction

Direct measurements of solar energetic particles associated with solar flares and coronal mass ejections have only been available since the 1960s. The spectral shape and fluence of solar proton events (SPEs) prior to the launch of cosmic ray detectors aboard satellites can be inferred using records from ground-based ionization chambers and neutron monitors, available since the 1930s, or visual observations of sunspots and solar flares extending back to the 1600s. There has also been considerable effort to identify signatures of solar events within paleoclimate records in order to quantify the historical frequency and intensity of events, necessary to predict future solar activity and to study the connection between solar variability and climate.

The use of nitrate ion measurements in Antarctic and Greenland ice cores has been proposed as a method to estimate the intensity, frequency, and probability of individual solar proton events [e.g, *Zeller and Dreschhoff*, 1995; *McCracken et al.,* 2001; *Kepko et al.,* 2009], to forecast future solar activity [e.g., *Kim et al.,* 2009; *Barnard et al.,* 2011; *Riley,* 2012], and to determine upper limits for the fluence and energies of solar cosmic rays [e.g., *Shea et al.,* 2006; *Townsend et al.,* 2006; *Shea and Smart,* 2012; *Miroshnichenko and Nymmik,* 2014]. However, the use of nitrate as a proxy has proven controversial as a result of: 1) the inability of models to reproduce enough atmospheric odd nitrogen or the transport mechanisms required to explain nitrate ion peaks in surface snow and ice [e.g., *Legrand et al.,* 1989; *Legrand and Kirchner,* 1990; *Duderstadt et al.,* 2014], 2) chemical correlations within ice cores pointing to alternative tropospheric sources [e.g., *Legrand and Mayewski,* 1997; *Wolff et al.,* 2008; 2012], and 3) the recognition of post-depositional processes involving nitrate ions during sequestration within ice cores [e.g., *Dibb and Whitlow,* 1996; *Dibb and Jaffrezo, 1997; Rothlisberger et al.,* 2002]. For these reasons, *Schrijver et al.* [2012] exclude nitrate as a proxy for SPEs in their analysis of the frequency of extreme solar events. Nonetheless, the use of nitrate as a proxy for SPEs persists, exemplified by the *Smart et al.* [2014] counterargument that only very high resolution nitrate ion measurements can capture SPEs, such as the analysis of the Greenland Ice Sheet Project (GISP2-H) and Boston University (BU) ice cores from Summit, Greenland.

Solar protons enhance odd nitrogen ($NO_y =$ $N+NO_2+NO_3+2N_2O_5+HNO_3+HO_2NO_2+ClONO_2+BrONO_2$) and odd hydrogen ($HO_x =$ $H+OH+HO_2$) in the polar upper stratosphere and mesosphere through the ionization and dissociation of molecular $N_2$ and $O_2$ [e.g., *Crutzen et al.,* 1975; *Jackman et al.,*1980; *Solomon et al.,* 1981]. Strong downward transport within the winter polar vortex brings enhanced $NO_y$ to the lower stratosphere; the $NO_y$ is potentially deposited at the surface as nitrate ions through the gravitational settling of polar stratospheric cloud particles or through stratosphere-troposphere exchange followed by wet or dry deposition. Satellite observations of $NO_2$ have confirmed the SPE enhancement of $NO_y$ and its subsequent descent [e.g., *López-Puertas et al.,* 2005; *Randall et al.,* 2005; *Jackman et al.,* 2009; *Randall et al.,* 2009]. However, modeling studies have been unable to produce sufficient enhancement of odd nitrogen from SPEs in the upper atmosphere to account for spikes seen in surface snow and ice [e.g., *Jackman et al.,* 1990; *Vitt et al.,* 2000; *Calisto et al.,* 2012; *Duderstadt et al.,* 2014]. In addition, given its long stratospheric residence time, diabatic recirculation, and dilution to lower latitudes, SPE-enhanced $NO_y$ in the mesosphere and stratosphere from individual events is unlikely to be discernible at the surface [e.g., *Legrand et al.,* 1989; *Legrand and Kirchner,* 1990]. Most of these modeling studies consider energies limited to 1-300 MeV, corresponding to proton channels measured by the Energetic Particle Sensor (EPS) instruments on GOES satellites [e.g., *Jackman et al.,* 1980, 2005, 2008]. There have been several efforts to model the atmospheric impact of high energy solar protons by also including satellite High Energy Proton and Alpha Detector (HEPAD) instrument observations (300 to >700 MeV), inferences from ground level enhancements (GLEs), and extrapolations of satellite observations to higher energies with functional fits [e.g., *Rodger et al.,* 2008; *Calisto et al.,* 2012, 2013; *Kovaltsov and Usoskin.,* 2014]. These studies focus on the effect of large SPEs on stratospheric ozone, temperature, and winds. The question remains whether higher energy SPEs, evidenced by GLEs from neutron monitors and muon detectors, produce enough $NO_y$ directly in the troposphere and lower stratosphere to account for spikes of nitrate ions at the surface.

Our work uses the Whole Atmosphere Community Climate Model (WACCM) to consider higher proton energies with the goal of estimating the fluence and spectra of solar protons that would be necessary to produce discernible nitrate ion spikes at the surface. We place our results in the context of the probability of occurrence of large SPEs derived from spacecraft, GLEs, and cosmogenic radionulide archives. How large do SPEs have to be in order to produce enhancements of $NO_y$ equivalent to the nitrate spikes attributed to SPEs in the GISP2-H and BU ice cores? Do these SPEs occur frequently enough for nitrate ion records to provide meaningful statistics of historical SPEs?

Section 2 of this manuscript discusses measurements of solar protons and strategies for extrapolating proton flux to higher energies. Section 3 describes the Whole

Atmosphere Community Climate Model (WACCM) and model scenarios. Section 4 presents model results in comparison with nitrate ion spikes in snow and ice. Section 5 discusses results in the context of the historical probability of solar extreme events. Section 6 summarizes conclusions regarding the efficacy of using nitrate peaks in ice cores to study historical SPEs.

## 2. The Fluence and Spectra of Solar Proton Events

A complete energy spectrum for SPEs can be constructed by extrapolating satellite proton flux spectra to relativistic energies, using ground-level neutron or muon measurements to infer solar proton flux, or a combination of both. For this work, we rely on highly documented SPEs as a basis for constructing hypothetically large events. We scale these SPEs in fluence and duration to produce atmospheric odd nitrogen enhancements of relative magnitudes similar to nitrate ion peaks attributed to SPEs in ice cores.

GOES Energetic Particle Sensors (EPS) measure seven logarithmically spaced proton differential energy bands ranging from 1 to ~500 MeV [*NASA,* 1996]. High Energy Proton and Alpha Particle Detectors [HEPAD] measure bands from 350 to 700 MeV, with an additional integral energy channel above 700 MeV. Based on the discussions of *Smart and Shea* [1999] and [*NASA,* 2006], we do not include channel P7 of EPS in our spectrum fits because the P7 spectral width (165-500 MeV) results in significant uncertainty regarding the choice of a midpoint energy. For the HEPAD channel P11, measuring the integral flux of energies greater than 700 MeV, we assume a midpoint energy of 1000 MeV following recommendations by *Smart and Shea* [1999].

While *Smart and Shea* [1999] also outline a strategy for compensating for side and rear penetration of high energy particles to lower energy channels, we have chosen not to include these corrections as they appear to be minimal and limited to specific events [*Mottl and Nymmik,* 2007] and are not consistent with our efforts to estimate the maximal potential enhancement of odd nitrogen from SPEs.

Optimization methods use ground-level measurements from neutron monitors to infer solar energetic particle spectra at the top of the atmosphere. We rely on the proton flux from these studies to model the 23 February 1956 SPE, an event widely used to estimate the strength of the 1859 Carrington white flare event as well as to predict future "worst case scenarios" [e.g., *Feynman et al.,* 1990; *Shea and Smart,* 1990; *Webber et al.,* 2007]. Similar methods are also used to identify functions for extrapolating satellite observations to higher energies.

Past studies have used a variety of spectral shapes to construct proton flux energy distributions from measurements. Prior WACCM experiments use exponential fits to interpolate between proton flux observations ranging from 1-300 MeV [*Jackman et al.,* 2008, 2009, 2011; *Funke et al.,* 2011; *Duderstadt et al.,* 2014]. Other studies rely on Bessel functions [*McGuire and von Rosenvinge,* 1984], Weibull functions [e.g. *Xapsos et al.,* 2000; *Kim et al.,* 2009] and power-laws with exponential rollover [e.g., *Ellison and Ramaty,* 1985; *Mazur et al.,* 1992; *Mewaldt et al.,* 2005]. The double-power-law function proposed by *Band et al.* [1993] appears to provide the best fit to events during the satellite era [e.g., *Mewaldt et al.,* 2005, 2012; *Tylka et al.,* 2006; *Atwell et al.,* 2011] and is particularly well-suited to energies above 100 MeV [e.g., *Mottl et al.,* 2001; *Mewaldt et al.,* 2012]. This Band function results in distinct power law parameters for low and high energies with a break (a.k.a., "knee," "bend point," or "spectrum turnover") at energies close to 30 MeV [e.g., *Mewaldt et al.,* 2005; *Wang and Wang,* 2006, *Mewaldt et al.,* 2012].

We interpolate observed SPE spectra to 60 logarithmically-based energy levels between 1 and 700 MeV, using an exponential fit between each measured EPS and HEPAD energy level. We then extrapolate to energies beyond 700 MeV by fitting a power law to channels above P5 (40-80 MeV), beyond the break energies typically seen in double power law spectra [*Mewaldt et al.,* 2012]. Questions remain concerning the shape of the proton spectra at relativistic energies [e.g. *Gopalswamy et al., 2012*; *Mewaldt et al.,* 2012; *Miroshnichenko et al.,* 2013]. For example, *Mewaldt et al.* [2012] identify a second spectral roll-over around 500 MeV based on GLE data. *Miroshnichenko and Nymmik* [2014] also point out overestimates to the high-energy tails of early reconstructions from GLEs such as the 23 Feb 1956 event. Direct measurements of protons with rigidities ranging from 1 GV to 1.2 TV by the Payload for Antimatter Matter Exploration and Light-nuclei Astrophysics (PAMELA) mission [*Adriani et al.,* 2011] and 1 GV to 1.8 TV by the Alpha Magnetic Spectrometer (AMS) on the International Space Station [*Aguilar et al.,* 2015] will allow more accurate estimates of higher energies. The spectral indices for the high-energy tails of hard spectrum hypothetical SPEs in our WACCM simulations are consistent with preliminary AMS data from 2011 to 2013 [*Aguilar et al.,* 2015]. On the basis of current knowledge, our high-energy tail extrapolations are consequently considered an upper bound, emphasizing the goal of estimating the maximum amount of odd nitrogen that could be produced in the lower atmosphere during an SPE.

The SPE proton flux reaching the top of the atmosphere evolves with time [e.g., *Wang et al.,* 2009; *Vashenyuk et al.,* 2011; *Miroshnichenko et al.,* 2013], usually beginning with a "prompt" component that is impulsive, has anisotropic pitch angles, and exhibits a "hard" (high-energy) exponential energy spectrum. A subsequent "delayed" component has a smoother time profile, is close to isotropic, and has a "soft" (low-energy) power law spectrum. The majority of high-energy particles are found during the first 20 hours of the event [*Li et al.,* 2003]. In this work, we capture this time evolution in spectral shape by basing our ion pair calculations on 30-minute average differential energy flux observations, in contrast to the daily averages used in prior WACCM simulations.

Our construction of hypothetical extreme SPEs from known SPEs assumes a uniform scaling across all energies and preserves the time evolution of the observed spectral shape. This strategy is based on the study of *Crosby* [2009], arguing that solar extreme events can be viewed as part of a global distribution originating from similar solar and heliospheric mechanisms as opposed to outliers with peculiar characteristics. However, in the absence of direct measurements of the characteristics of SPEs at the high fluences discussed in this paper, the spectral shape and evolution of our hypothetical constructions remain uncertain.

## 3. WACCM Model

The Whole Atmosphere Community Climate Model (WACCM) is a three dimensional Earth system model specifically designed to study the impact of upper atmospheric processes on global climate. WACCM is part of the Community Earth

System Model (CESM) at the National Center for Atmospheric Research (NCAR) and combines the Community Atmosphere Model (CAM5), the Thermosphere-Ionosphere-Mesosphere-Electrodynamics General Circulation Model (TIME-GCM), and the Model for OZone and Related chemical Tracers (MOZART) [*Garcia et al.,* 2007; *Kinnison et al.,* 2007; *Emmons et al.,* 2010; *Marsh et al.,* 2013; *Neale et al.,* 2013]. By resolving the entire stratosphere and mesosphere, WACCM is ideally suited for modeling SPE production of $HO_x$ and $NO_x$ in the upper atmosphere [e.g., *Jackman et al.,* 2008, 2009, 2011; *Funke et al.,* 2011; *Duderstadt et al.,* 2014].

For primary particles with energies of 1-300 MeV, ion pair production rates as a function of pressure are calculated using the parameterization of *Jackman et al.* [1980, 2005, 2008] based on the range data of *Sternheimer* [1959]. Higher energies use ionization yield functions from *Usoskin et al.* [2010] calculated using the CRAC:CRII (Cosmic Ray-induced Atmospheric Cascade: Application for Cosmic Ray Induced Ionization) particle transport model. The Monte Carlo methods of the CRAC:CRII reaction cascade resolve species and processes below 1 g·cm$^{-2}$ ($\sim$ 50 km) [*Usoskin et al.,* 2004; *Usoskin and Kovaltsov, 2006; Usoskin et al.,* 2010] and account for the effects of nuclear processes and secondary particles.

*Jackman et al.* [1980, 2005] describe WACCM parameterizations for $HO_x$ and $NO_x$ production. WACCM relies on a lookup table to calculate the production of $HO_x$, resulting from the ionization of $N_2$ and $O_2$ followed by water cluster ion formation [*Solomon et al.*, 1981]. Regarding $NO_x$ production in WACCM, the ionization and dissociation of $N_2$ and $O_2$ produce 1.25 N per ion pair [*Porter et al.,* 1976], subsequently partitioned into 45% ground state $N(^4S)$ and 55% excited state $N(^2D)$, with the excited state determining NO production [*Rusch et al.,* 1981]. These simulations assume uniform $HO_x$ and $NO_x$ production at geomagnetic latitudes greater than 60 degrees.

Our WACCM experiments have a horizontal resolution of 1.9 degrees latitude by 2.5 degrees longitude with 88 vertical layers extending to ~140 km. These simulations use the Specified Dynamics (SD) version of WACCM. Meteorological fields in the model from 0-40 km are nudged to data from the NASA Global Modeling and Assimilation Office (GMAO) Modern Era Retrospective Analysis for Research and Application (MERRA) meteorological reanalysis [*Rienecker et al.,* 2011]. There is a linear transition in the nudging from 40 to 50 km, above which the model is free-running. To study the sensitivity of nitrate production to solar proton spectral shape and fluence, we reproduce several known SPEs assuming a start date of 18 December 2004. The Arctic winter of 2004-2005 was chosen because it was unusually cold, with strong diabatic descent and limited mixing to lower latitudes during December and January [*Manney et al.,* 2006]. Any stratospheric $NO_y$ produced by an SPE under these meteorological conditions would remain well sequestered within the polar vortex with relatively unmixed descent for at least six weeks. SD-WACCM simulations for the 2004-2005 polar winter also show good correlation with satellite measurements of ozone and related constituents [*Brakebusch et al.,* 2013]. While *Brakebusch et al.* [2013] conclude that SD-WACCM overestimates mixing across the vortex edge in the latter half of the season, the model captures vertical transport well within the low-mid stratosphere, especially during December and January. This study does not address the sensitivity to dynamical conditions and is limited to the Northern Hemisphere. Although the 2004-2005 Arctic winter was cold, it is likely that conditions even more conducive to descent occurred in the past. However, even if all the $NO_x$ produced by the hypothetical SPEs was sequestered in the polar vortex, the conclusions drawn here would not change.

WACCM includes the simulation of polar stratospheric clouds (PSCs). Although not as significant in the Arctic as the Antarctic, PSCs play a role in atmospheric chemistry during the 2004-2005 winter as a result of the cold, stable vortex early in the winter [*Brakebusch et al.,* 2013]. WACCM addresses heterogeneous reactions on sulfate aerosols, nitric acid trihydrate (NAT), supercooled ternary solution (STS), and water ice associated with PSCs [*Kinnison et al.,* 2007]. Including simulations of PSCs is important for studying nitrate deposition due to the uptake of nitrate on PSCs followed by gravitational settling of PSCs to the troposphere. However, it is important to keep in mind that PSCs are infrequent in the Arctic, and occur only inside the polar vortex during cold winters, with average occurrence frequencies of less than 6% [*Fromm et al.,* 2003].

## 4. Results

### 4.1. WACCM Experiment Setup

Prior global model simulations of SPEs have failed to produce enough $NO_y$ in the upper atmosphere to account for nitrate ion spikes observed in snow and ice, generally showing enhancements of only 10-20% in the lower stratosphere during the months following large SPEs [e.g., *Vitt et al.,* 2000; *Jackman et al.,* 2008, 2009; *Duderstadt et al.,* 2014]. *Duderstadt et al.* [2014] conducted WACCM simulations for the 9 November 2000 SPE, the sixth largest event in the past 50 years in terms of $NO_y$ production. The authors chose this event because of the availability of concurrent, year-round, daily snow surface measurements at Summit, Greenland, including a comprehensive suite of ion measurements capable of distinguishing probable tropospheric sources of nitrate from possible SPE influences. While WACCM calculates a thin ~5 km layer of 5-10 ppbv SPE-enhanced $NO_y$ at 25-30 km in the months following the 9 Nov 2000 event, the thick background pool of 10-15 ppbv $NO_y$ in the lower stratosphere results in a total column atmospheric enhancement of less than 5%, not enough to explain the twofold to fivefold spikes in nitrate observed in surface snow. Justification for the use of $NO_y$ column densities and polar vortex averages is described in detail in *Duderstadt et al.* [2014]. The 9 November 2000 event, however, was not the highest fluence event in past decades nor did it produce GLEs at the surface. In our current study, we expand upon this prior research by considering SPEs with larger fluences and harder ("higher energy") spectra, noting that daily snow measurement comparisons are not available for these events.

Table 1 compares several of the largest SPEs occurring since the 1950s (derived from *Reedy et al.* [2006], *Shea et al.* [2006], *Webber et al.* [2007], *Jackman et al.* [2008, 2009, 2014], *McCracken et al.* [2012], and *Kovaltsov et al.* [2014]). The characteristics of events prior to the satellite era are inferred from ground level measurements of secondary particles [e.g., *Feynman et al.,* 1990; *Shea and Smart,* 1990]. Table 2 summarizes the modeling experiments presented in this paper. Our first set of WACCM experiments (Section 4.2) compares several satellite era SPEs that have large fluences of protons with energy greater than 30 MeV (defined to be $F_{30}$), including the series of SPEs in October 1989, the 14 July 2000 "Bastille Day" event, the 9-11 Nov 2000 SPE, the 28-31 October 2003 "Halloween Storms," and the 23 July 2012 SPE that missed Earth but was observed by the STEREO-A spacecraft. A second set of modeling experiments (Section 4.3) evaluates odd nitrogen production by hard spectrum SPEs, specifically the events of 29

September 1989 and 20 January 2005 that resulted in the largest GLEs in the satellite era. Finally, we present simulations of the 4 August 1972 and 23 February 1956 SPEs (Section 4.4), comparing WACCM calculations with other studies striving to use "Carrington-like" conditions to study worst-case scenarios. We place our WACCM results in the context of nitrate ion spikes measured in the GISP2-H and BU ice cores from Summit, Greenland. For all simulations, we increase the flux and duration of events in order to determine the fluence and spectrum of an SPE necessary to produce enough odd nitrogen in the atmosphere to potentially be discernible as nitrate spikes at the surface.

Figures 1 and 2 show the differential energy spectra and ion pair production rates for several SPEs in this study. The vertical axis of Figure 1 represents the midpoint energy levels of the GOES proton channels (unequally spaced) and the black dashed lines denote 30 MeV and 200 MeV cutoffs corresponding to Table 1. The proton flux for the 23 Jul 2012 event uses hourly STEREO-A observations from 1-100 MeV extrapolated to 500 MeV based on the PREDICCS (Predictions of Radiation from REleASE, EMMREM, and Data Incorporating the CRaTER, COSTEP, and other SEP measurements) model [*Schwadron et al.,* 2012; *Joyce et al.,* 2013].

The 14 July 2000 event clearly has the largest $F_{30}$ fluence during peak intensity, resulting in high rates of ionization from 40-70 km. The extended duration of the October 1989 and October 2003 events explain their large total $F_{30}$ fluences as well, resulting in significant ionization throughout the upper stratosphere and mesosphere. The majority of ion pair production occurs above 20 km. Ionization below 20 km is evident during the 20 Jan 2005 SPE, and there are also high ionization rates at relatively low stratospheric altitudes during the 29 Sep 1989, 20 Oct 1989, and 14 Jul 2000 SPEs. These figures confirm the conclusion by *Usoskin et al.* [2011] and *Mironova and Usoskin* [2013] that ion pair production rates in the lower stratosphere are 1-2 orders of magnitude less than in the mid-stratosphere, even during hard spectrum events.

## 4.2. WACCM Simulations of High Fluence Events

Figure 3 presents $NO_y$ enhancements from several high fluence events ($F_{30}$) averaged over the Arctic polar vortex throughout winter. All events are placed beginning on 18 December 2004. The boundary of the meandering polar vortex is objectively determined by identifying grid points within the stratosphere where scaled potential vorticity (sPV) on isentropic surfaces exceeds $1.4 \times 10^{-4}$ s$^{-1}$ [*Dunkerton and Delisi,* 1986], as described in further detail in *Duderstadt et al.* [2014]. This figure illustrates $NO_y$ enhancements in the upper stratosphere and mesosphere during the events as well as diabatic descent within the cooling polar vortex throughout winter. This figure also shows background $NO_y$ (without SPEs) in units of mixing ratio and number density, highlighting the pool of $NO_y$ in the lower stratosphere formed from $N_2O$ oxidation. This background concentration of $NO_y$ can also be brought to the surface following tropopause folds (stratosphere-troposphere exchange) and denitrification on polar stratospheric cloud particles, suggesting that SPE enhancements of odd nitrogen would need to be much larger than this reservoir to be detected at the surface.

A comparison of total column density for simulations with and without SPEs provides a simple means of assessing SPE enhancements of odd nitrogen with respect to background. Figure 4 presents vortex-averaged total $NO_y$ column densities for these large fluence events. All SPE $NO_y$ enhancements remain below 15% relative to background levels, with localized enhancements (individual grid points within the polar vortex) below 20% (not shown).

We note that values of $NO_y$ presented in this paper include both gas and condensed phase $HNO_3$, allowing the study of the maximal potential for nitrate deposition from both gas-phase reactions and heterogeneous reactions on PSCs as well as the effects of the formation, settling, and evaporation of PSCs. Including the impact of PSCs on SPE enhanced $NO_y$ is particularly important during the 2004-2005 winter, with WACCM simulations suggesting NAT PSC formation beginning in early December from 20-30 km, gradually extending to 15-30 km by the end of January. Nitric acid from PSCs re-enters the gas phase below 15-20 km in these WACCM simulations, suggesting that gravitational settling of NAT PSCs and subsequent re-nitrification near or within the troposphere (generally extending in the Arctic from the surface to ~8-12 km) is unlikely. These results are consistent with re-nitrification observed during the Polar Aura Validation Experiment (PAVE) of January/February 2005 [*Dibb et al.,* 2006] as well as the previous WACCM analysis of SPEs and NAT PSCs in *Duderstadt et al.* [2014].

Most of the SPE enhancement of $NO_y$ occurs above 30 km. Figure 5 illustrates how relatively large increases in $NO_y$ mixing ratios in the upper atmosphere can lead to minimal enhancement at the surface, a result of the exponential decrease in atmospheric density with altitude. The left side of Figure 5 depicts $NO_y$ enhancements (mol/mol) during, two weeks after, and six weeks after each event, showing the descent of SPE-enhanced $NO_y$ with time. The corresponding values of cumulative column density (molecules cm$^{-2}$) demonstrate the minimal impact of SPEs on NO$_y$ in the troposphere, with enhancements at the surface remaining under 15%, consistent with Figure 4.

In the search for SPE-enhanced NO$_y$ discernible above background variability, we increase the fluence of the 19-27 October 1989 SPEs (the largest SPEs with respect to NO$_y$ production) by 10 times, 50 times, and 100 times. *Zeller and Dreschhoff* [1995] and *McCracken et al.* [2001] attribute a threefold peak in nitrate within the GISP2-H ice core to the 1859 Carrington Event (~300 ng/g compared to the summertime maximum of 100 ng/g, with seasonal variations ranging from 20-100 ng/g). The largest spikes in the GISP2-H core are five times larger than the seasonal cycle amplitude [*Zeller and Dreschhoff*, 1995; *Schrijver et al.*, 2012]. *Smart et al.* [2014] attribute a GISP2-H nitrate spike 2.5 times background (150 ng/g compared with summertime 60 ng/g) to the 23 Feb 1956 GLE. *Smart et al.* [2014] also identify a fivefold nitrate ion spike in the BU Summit ice core at a depth corresponding to the 25 July 1946 GLE (~500 ppb in comparison to adjacent ~100 ppb levels) and a threefold spike at a depth dated as corresponding to the 19 Nov 1949 GLE (~300 ppb in comparison to the summertime peak of ~100 ppb). Based on these examples, our target is to identify SPE fluences that result in twofold to fivefold SPE enhancements of NO$_y$.

Total column densities for the extreme SPEs based on the October 1989 events are presented in Figure 6. An event 50 times the October 1989 events would produce total column NO$_y$ enhancements two times background levels. An event 100 times the October 1989 events would result in threefold to fourfold enhancements during and following the SPEs, producing enough odd nitrogen throughout the atmospheric column to match the relative magnitude of nitrate spikes attributed to SPEs in the GISP2-H and BU ice cores by *Zeller and Dreschhoff* [1995], *Kepko et al.,* [2009] and *Smart et al.* [2014]. This total

column $NO_y$ enhancement lasts for several months, suggesting that any potential

association with enhanced nitrate deposition would likely occur over a period of months

rather than during a single snowfall as indicated by nitrate spikes observed in ice cores

We also consider the production of $NO_y$ directly in the troposphere and lower

stratosphere during the October 1989 events, especially given that several GLEs were

recorded in a short time period on 19, 22, and 24 October 1989. Figure 7a shows the

vortex-averaged column density of $NO_x$ ($NO + NO_2$) integrated throughout the

troposphere and lower stratosphere (0-15 km) for the October 1989 SPE scaled by 10

times, 50 times, and 100 times. We focus on $NO_x$ in order to capture the maximal

immediate SPE enhancement, noting that the lifetime of $NO_x$ with respect to oxidation to

$NO_y$ is days to weeks in the troposphere. We choose 0-15 km to remain below the peak

background pool of $NO_y$ as well as below the lowest altitude at which PSCs form during the simulation, recalling that PSCs are capable of removing $NO_x$ from the gas phase.

Figure 7b shows vortex-averaged 0-15 km column density of $NO_y$ from November

through March. It is clear that, even with the full spectrum of energies measured by the

GOES EPS and HEPAD instruments along with an extrapolated power law tail, there is

not enough $NO_x$ or $NO_y$ produced within 0-15 km to suggest a discernible nitrate

enhancement at the surface. (Corresponding tropospheric column density enhancements

of $NO_x$ and $NO_y$ from 0-10 km are less than 1%.)

Figure 7c and 7d show column $NO_y$ from 0-30 km and column gas phase $HNO_3$

plus NAT on PSCs from 15-30 km, addressing the potential for enhanced nitrate deposition through gravitational settling of PSCs forming in WACCM from 15-30 km during late December through January. The Oct 89 x100 SPE results in up to a 70% enhancement in 0-30 km $NO_y$ during times when PSCs are present in WACCM (recall that $NO_y$ includes nitrate as NAT on PSCs) and a 70% increase in 15-30 km $HNO_3$ + NAT on PSCs.

The short-lived tropospheric $NO_x$ signal is the most reliable indicator for sharp spikes in nitrate deposition, representing direct production in the troposphere. The enhancements of $NO_y$ at all higher levels in the atmosphere lack the characteristics necessary to produce a sharp spike as they last for months as a result of longer lifetimes and slow vertical descent. This behavior includes the gradual enhancement of 0-15 km $NO_y$ that is most likely the result of downward transport from the stratosphere.

### 4.3. WACCM Simulations of Hard Spectrum Events

From the analysis in the previous section, we conclude that events with the highest $F_{30}$ fluence over the past several solar cycles do not produce $NO_y$ enhancements of similar magnitude to the peaks observed in ice cores, either throughout the atmospheric column or directly in the lower atmosphere. However, the largest fluence events do not necessarily correspond with the events exhibiting highest energies (hardest spectra). Higher energy protons are required to produce spallation products that can penetrate and ionize the lower atmosphere. In this section, we investigate the potential for direct production of $NO_x$ at altitudes below the natural background pool of $NO_y$. This enhanced $NO_x$ would then need to be converted to $HNO_3$ and deposited to the surface, either directly through wet deposition in the troposphere or via tropopause folds or gravitational settling of polar stratospheric cloud particles in the lower stratosphere.

We base our simulations of hard events on the 29 September 1989 and 20 January 2005 SPEs, with 20 Jan 2005 having the second largest GLE and 29 Sep 1989 having the fourth largest GLE since the 1940s [*Wang et al.,* 2009]. We uniformly scale up the proton fluxes of the 29 Sep 1989 and 20 Jan 2005 events until we reach sufficient $NO_y$ enhancements to potentially be observed at the surface. We assume an isotropic proton flux for ion pair calculations but acknowledge observational evidence of a short-lived anisotropy for the 20 Jan 2005 event [*Bazilevskaya et al.,* 2008].

Figure 8 shows the extrapolated spectral fits to GOES proton data during the prompt components of the 29 Sep 1989 and 20 Feb 2005 SPEs. The power law spectral index for the tail of the 20 Jan 2005 SPE (-2.4 to -2.6) is slightly larger than the *Mewaldt et al.* [2012] fluence spectral index of -2.14 (+/- 0.06). *Mewaldt et al.* [2012] suggest that spectral indices of at least -2.8 are necessary to produce GLEs. *Tylka and Dietrich* (2009) argue that the spectrum steepens for energies greater than ~500 MeV (rigidities greater than ~1 GV), with average integral spectral slopes in rigidity (momentum/charge) of -5 to -7. The extrapolations used in this study consequently represent overestimates of spectral flux, consistent with efforts to determine the minimal characteristics of an SPE necessary to enhance nitrate at the surface.

A more robust estimate of ionization rates and $NO_y$ production requires the addition of galactic cosmic rays (GCRs). While including these particles will increase background ion pair production in the lower atmosphere, it tends to decrease ionization during SPEs as a result of perturbations to the heliospheric magnetic field and/or interplanetary shock; this is known as a Forbush decrease. *Usoskin et al.* [2009], for example, include GCRs in model calculations for the 20 Jan 2005 SPE, concluding that this event enhances ionization only in the middle and upper polar atmosphere, with a Forbush decrease suppressing ionization at lower altitudes and latitudes. Because we want to determine the minimal solar proton flux necessary to achieve maximal enhancement of $NO_y$, we have chosen not to include GCRs in these hard spectra simulations.

We perform a series of case studies increasing the fluxes of the 29 Sep 1989 and 20 Jan 2005 SPEs as well as extending the duration of the events. Figure 9 presents WACCM calculations of vortex-averaged 0-15 km $NO_x$ and $NO_y$, 0-30 km $NO_y$, and 15-30 km $HNO_3$ for the 20 Jan 2005 SPE, with an analysis rationale similar to Figure 7 described in section 4.2. The 20 Jan 2005 10dx100 SPE results in the greatest direct enhancement of $NO_x$ in the troposphere and lower stratosphere, peaking at 10% during the event (Figure 9a). $NO_y$ SPE enhancements continue to increase throughout the winter, consistent with descending $NO_y$ from the upper stratosphere, peaking at 40% for the 20 Jan 2005 x1000 SPE (Figure 9b). Corresponding peak enhancements for the 29 Sep 1989 SPE simulations are ~2% $NO_x$ and ~25% $NO_y$ for 0-15 km column densities during the 29 Sep 1989 10dx100 SPE (not shown). None of the hypothetical SPEs are capable of producing enhancements of $NO_x$ greater than 1% directly in the troposphere (~0-10 km).

Figures 9c and 9d suggest the potential for two-fold enhancements of nitrate deposition from the region impacted by nitrate deposition through PSCs, with WACCM calculations for the hypothetical 20 Jan 2005 x1000 and 20 Jan 2005 10dx100 SPEs producing ~130% enhancements of 0-30 km column $NO_y$ and ~150% enhancements of 15-30 km column $HNO_3$ + NAT on PSCs from December through January. Corresponding enhancements for the 29 Sep 1989 simulations are 60% for 0-30 km column $NO_y$ and 70% for 15-30 km column $HNO_3$ + NAT on PSCs, respectively (not shown).

Figure 10 presents the $NO_y$ total column density for these simulations. An event 1000 times the 29 Sep 1989 SPE produces a fourfold total column increase in $NO_y$, which decreases to threefold in the weeks following the event. An event 1000 times the 20 Jan 2005 SPE produces a fourfold to fivefold increase in $NO_y$ during and following the event. The 10dx100 scenario, which has an equivalent total fluence to the x1000 event, produces a threefold to fourfold increase.

### 4.4. WACCM Simulations of "Carrington-like" Events

Estimates for the September 1859 Carrington Event are often used to represent worst-case scenarios in preparation for future solar storms. Most of these studies are based on the $1.9x10^{10}$ $cm^{-2}$ $F_{30}$ proton fluence derived from GISP2-H nitrate measurements [*McCracken et al.,* 2001] and aim to quantify the impacts of worst-case scenarios on atmospheric chemistry, temperature, dynamics, and radiation doses. *McCracken et al.* [2001] derive this estimate from a nitrate anomaly in the GISP2-H core dated to the year 1859 by *Zeller and Dreschhoff* [1995]. Following the *Wolff et al.* [2012] study comparing the GISP2-H core to more robustly dated high-resolution cores, this nitrate spike is likely associated with a 1863 biomass burning event observed in other Greenland cores. This estimate consequently no longer has any observational basis. Nonetheless, we conduct WACCM simulations to determine if this magnitude of $F_{30}$ fluence could produce threefold enhancements in $NO_y$ similar in relative magnitude to the GISP2-H nitrate ion spike. While this approach is somewhat circular, it provides a simple means to determine if the underlying assumption of the $1.9 \times 10^{10}$ cm$^{-2}$ $F_{30}$ used in prior studies to estimate the fluence for a worst-case scenario is well founded.

There are many examples of prior Carrington-like event simulations. *Townsend et al.* [2003, 2006] use a Weibull fit for the 4 Aug 1972, Aug-Sep-Oct 1989, and 23 Mar 1991 SPEs scaled to the $1.9 \times 10^{10}$ cm$^{-2}$ $F_{30}$ fluence to represent the Carrington Event. *Rodger et al.* [2008] consider Weibull fits for the 4 Aug 1972 and 23 Mar 1991 SPEs scaled to the $1.9 \times 10^{10}$ cm$^{-2}$ $F_{30}$ fluence as well, arguing that soft energy spectra are most representative of the Carrington event due to the lack of signature for the event in $^{10}$Be ice core data. *Miroshnichenko and Nymmik* [2014] take the average spectrum from five SPEs with the largest $F_{30}$ fluence from solar cycle 23 (8 Nov 2000, 24 Sep 2001, 4 Nov 2001, 28 Oct 2003, and 17 Jan 2005) to obtain a differential spectrum representative of the Carrington Event, scaling it once again to the $1.9 \times 10^{10}$ cm$^{-2}$ $F_{30}$ fluence. *Calisto et al.* [2012] similarly scale the 4 Aug 1972 event to the $F_{30}$ fluence of $1.9 \times 10^{10}$ cm$^{-2}$ $F_{30}$, using the time profile reconstructed by *Smart et al.* [2006a] and the three-dimensional Chemistry Climate Model SOCOL v2.0 to study atmospheric impacts. *Calisto et al.* [2013] improve on this effort to study a worst-case scenario by basing a Carrington-like event on the hard 23 Feb 1956 SPE and adding GCRs. The authors calculate ionization rates for high energy protons using the CRAC:CRII particle transport model. *Calisto et al.* [2013] find statistically significant SPE influences not only on upper atmospheric $HO_x$, $NO_x$, and $O_3$ but also on zonal winds and surface temperatures. The authors calculate up to a 40% increase in $HNO_3$ deposition during the month following the Carrington-like 1956 simulations in regions over the South Pole but note a less pronounced effect in the Arctic.

We conduct similar Carrington-like simulations based on the Aug 1972 and Feb 1956 proton fluxes, using the ionization profiles of *Usoskin et al.* [2010] and *Calisto et al.* [2013]. Figure 11 shows the ionization rates used as input for WACCM. These simulations include ion pair production rates from GCRs as described in *Calisto et al.* [2013]. Figure 12 presents a comparison of the peak ionization rate (IR) profiles used in WACCM for the Carrington-like Soft (1972) and Hard (1956) simulations along with the Oct 1989, Jan 2012, and Mar 2012 rates calculated using 1-300 MeV protons [*Jackman et al.,* 2014]. This figure also shows the Aug 1972 ionization rates from *Rodger et al.* [2008] and *Calisto et al.* [2012] and annual mean GCR rates by *Mertens et al.* [2013]. It is clear that the Hard Carrington-like event based on the 23 Feb 1956 SPE results in the greatest levels of total ionization, with significantly more ionization in the lower atmosphere than background GCRs (one to two orders of magnitude).

As a comparison with the previous model runs, Figure 13 presents the vortex-averaged $NO_y$ total column density for both the Carrington-like Soft (Soft Car 1972) and Carrington-like Hard (Hard Car 1956) events, where the reconstructed September 1 Carrington peak is placed on 13 December 2004. The Soft Car 1972 event results in a 5% total column increase in $NO_y$ while the Hard Car 1956 event produces a 20% total column increase in $NO_y$. This result is not surprising considering the greater atmospheric ionization rates of the Hard Car 1956 event at lower altitudes, where atmospheric densities are much higher. These Carrington-like events do not produce enough odd nitrogen within the atmospheric column to account for the threefold spike in nitrate ions attributed to the Carrington Event in the GISP2-H ice core.

As a result of the Hard Car 56 simulated event, WACCM calculates a ~3% maximal increase in $NO_x$ vortex-averaged column densities for 0-10 km and ~15% for 0-15 km during the event, with < 10% enhancements of 0-15 km column density throughout winter. Enough $NO_x$ is produced in the atmosphere to result in a 1% reduction of $O_3$ at 20 km and 10-20% reduction of $O_3$ from 30 km to 40 km in the weeks following the event, consistent with the results of *Calisto et al.* [2013]. However, even with rapid oxidation of $NO_x$ to $HNO_3$ and strong stratosphere-tropospheric exchange, this enhancement would not be enough to account for the nitrate peaks observed in ice cores. The potential impact of gravitational settling of nitrate on PSCs is also minimal, with enhancements of 10% for 0-30 km column densities of $NO_y$ (including NAT on PSCs) and 15% for 15-30 km column densities of $HNO_3$ + NAT on PSCs.

It is unlikely that a $1.9x10^{10}$ cm$^{-2}$ $F_{30}$ fluence could produce a threefold enhancement of $NO_y$ in the lower atmosphere similar to the nitrate spike in the GISP2-H ice core attributed to the Carrington Event. However, as we discuss in the next section, satellite and cosmogenic radionuclide studies show that a $F_{30}$ fluence of this magnitude is within the occurrence probability of a millennium timescale, supporting prior studies involving worst-case scenarios based on this fluence.

**5. Discussion**

Our WACCM experiments indicate that events 50 times and 100 times larger than the 1989 October series of events would be required to increase $NO_y$ by twofold and threefold to fourfold, respectively, throughout the atmosphere. An event three orders of magnitude larger than the hard spectrum 20 Jan 2005 and 29 Sep 1989 events would be necessary to produce a short-lived twofold to threefold increase of $NO_x$ at altitudes below the natural pool of stratospheric $NO_y$. These enhancements are similar in relative magnitude to the twofold to fivefold nitrate spikes measured in Greenland ice cores and attributed to SPEs.

By placing these hypothetical SPEs in the context of probability distributions and upper limit estimates, we can assess the frequency that such events might occur and consequently the potential usefulness of nitrate as a proxy for individual SPEs. It is important to keep in mind that we are considering only the contribution of odd nitrogen enhancement in the atmosphere. There are many additional arguments challenging the use of nitrate as a proxy for SPEs that are beyond the scope of this paper, including stratospheric residence time, vertical and latitudinal mixing, alternative sources of nitrate, sequestration of nitrate within buried ice, and post-depositional processing.

Figure 14 presents total integrated fluences for the hypothetical Oct 89 x50 and Oct 89 x100 SPEs overlaid onto fluence summaries from *Webber et al.* [2007] and *Beer et al.* [2012] for events from 1956 to 2005. This figure illustrates how extreme an SPE would have to be in order to produce two to four times the amount of total column $NO_y$. These hypothetical events are well above the integral fluences calculated for measured events. The red star represents the $F_{30}$ estimate of $1.9x10^{10}$ cm$^{-2}$ for the Carrington event from *McCracken et al.* [2001] used in section 4.4, well below the magnitude of the Oct 1989 x50 SPE needed to produce a doubling of $NO_y$ throughout the atmosphere.

Several studies calculate the probability of occurrence of SPEs using satellite measurements, GLEs, and terrestrial and lunar cosmogenic isotopes. The slope of the cumulative probability of occurrence becomes steeper at larger fluences, resulting in a broken power law distribution [*Smart et al., 2006b*]. *Usoskin and Kovaltsov* [2012] analyze cosmogenic isotopes [14]C and [10]Be in terrestrial archives to determine the probability of occurrence of SPEs with respect to $F_{30}$. They calculate conservative $F_{30}$ limits of $1 \times 10^{10}$ cm$^{-2}$, $2$-$3 \times 10^{10}$ cm$^{-2}$, and $5 \times 10^{10}$ cm$^{-2}$ occurring on the order of 100 years, 1000 years, and 10,000 years, respectively. These levels are plotted as black circles on Figure 14 and are below the $F_{30}$ fluences of our hypothetical scenarios. *Kovaltsov and Usoskin* [2014] arrive at a cumulative occurrence probability distribution function based on 60 years of direct fluence measurements, terrestrial [14]C and [10]Be records during the Holocene, and cosmogenic radionuclides in lunar rocks. The authors confirm that SPEs with $F_{30}$ fluences greater than $10^{11}$ protons cm$^{-2}$ yr$^{-1}$ are not expected on a millennium timescale, and no events greater than $5 \times 10^{11}$ protons yr$^{-1}$ are expected to have occurred during the past 10,000 years. The probabilties at higher fluences of *Usoskin and Kovaltsov* [2012] and *Kovaltsov et al.* [2014] rely on the identification of [14]C and [10]Be peaks in ice core archives, with significant uncertainty regarding the contributions from Earth system effects and local meteorology. Nonetheless, cosmogenic radionuclides remain the current best estimate. *Miroshnechenko and Nymmik* [2014] evaluate the fluence of events measured by satellites during recent solar cycles combined with historical sunspot data to conclude that an event with $F_{30}$ greater than $6 \times 10^{10}$ cm$^{-2}$ is only likely to occur once every $2.6 \times 10^5$ years.

The Oct 1989 series of events (19-27 Oct) has a total $F_{30}$ fluence of $\sim 4 \times 10^9$ cm$^{-2}$ (Table 1). An event 50 times the Oct 1989 events, necessary to double polar total column odd nitrogen, would consequently have a $F_{30}$ fluence of $\sim 2 \times 10^{11}$ cm$^{-2}$, which is well beyond the Holocene probability estimates from *Usoskin and Kovaltsov* [2012] and *Miroshinenko and Nymmik* [2014]. An event 100 times the Oct 1989 event would have a $F_{30}$ fluence of ~$4 \times 10^{11}$ cm$^{-2}$, approaching the Holocene upper limit indicated by *Kovaltsov and Usoskin* [2014]. *Usoskin et al.* [2013] estimate that an extremely large SPE capable of producing the spikes observed in 774-775 AD cosmogenic radionuclide archives [*Miyake et al.,* 2012, 2015] would be 25-50 times stronger than the Feb 1956 SPE which had a $F_{30}$ fluence of ~ $8 \times 10^{10}$ cm$^{-2}$ as estimated by *Cliver et al.* [2014] (purple hexagon in Figure 14). *Thomas et al.* [2013] assert that a smaller event, an order of magnitude greater than the October 89 SPEs, would explain the 774-775 spike. Both estimated fluences are less than the amount required by WACCM to produce a doubling of odd nitrogen in the atmospheric column, implying that signatures of a 774-775 AD event are not expected in ice core nitrate records. In addition, *Cliver et al.* [2014] question an event of this size, suggesting that it would require an active region on the Sun (sunspot size) 2.5 times larger than has been recorded, further challenging the likelihood of the extreme hypothetical Oct 89 x50 and Oct 89 x100 events. This argument is consistent with the *Aulanier et al.* [2012] estimate for an upper limit to solar flare energies equivalent to six times the 4 Nov 2003 flare. The implications of this analysis are that SPE-enhanced odd nitrogen in the mesosphere and upper stratosphere is insufficient to produce discernible nitrate spikes at the surface for any meaningful analysis of individual SPE statistics during the Holocene.

Next, we assess the probability of hard spectrum SPEs producing odd nitrogen directly in the lower atmosphere by analyzing the total integral fluence above 200 MeV (defined to be $F_{200}$) as well as the instantaneous flux of protons during peak intensity of our hypothetical hard spectrum SPEs. Figure 15 is adapted from *Wang et al.* [2009] and summarizes instantaneous solar proton integral flux as a function of energy during times of peak intensity for several measured events. Overlaid onto this figure are the integral flux spectra during the prompt component of the 29 Sep 1989 SPE and 20 Jan 2005 SPE used in our WACCM simulations along with the corresponding 1000-fold hypothetical events. *Wang et al.* [2009] calculate spectra using neutron monitor data, ionization chamber data, and GOES and Meteor direct satellite measurements. Our simulations use power law fits from 30-minute average GOES data, resulting in larger flux at relativistic energies as discussed in Section 2.

The black dashed line in Figure 15 represents a theoretical Upper Limit Spectrum [*Miroshnechenko and Nymmik*, 2014], calculated by averaging the time-of-maximum intensity spectra of the five largest $F_{30}$ SPEs of solar cycle 23 and scaling to a fluence an order of magnitude larger than the $1.9x10^{10}$ $cm^{-2}$ *McCracken et al.* [2001] estimation for the Carrington Event. We note that although the fluence of this Upper Limit Spectrum is determined using a nitrate source that has been challenged by *Wolff et al.* [2012] the shape of the spectrum is determined from direct measurements and is remarkably consistent with the shape of our hypothetical events from 30 MeV to 10 GeV. The peak fluxes for the hypothetical 29 Sep 89 x1000 and 20 Jan 05 x1000 events exceed all other events during the satellite era as well as the *Miroshnechenko and Nymmik* [2014] Upper Limit Spectrum.

Figure 16 shows probability of occurrence distribution functions for $F_{30}$ and $F_{200}$ from *Usoskin and Kovaltsov* [2012] and *Kovaltsov et al.* [2014] based on satellite data and cosmogenic radionuclides. The vertical gray lines indicate the fluences of the hypothetical events in the WACCM simulations. The 29 Sep 89 x100 and 20 Jan 05 x100 hypothetical SPEs, producing only a 15-40% enhancement of $NO_x$ in the lower stratosphere, occur on the order of once every 100-1000 years and would likely not be distinguishable from seasonal variability. *Usoskin and Kovaltsov* [2012] use [10]Be data to identify 19 candidates throughout the Holocene having 10-30 times the fluence of the 23 Feb 1956 SPE. These events, with $F_{200}$ the order of $10^9$ cm$^{-2}$ and $F_{30}$ on the order of $10^{10}$ cm$^{-2}$, would be well below the limits needed to produce a doubling of odd nitrogen in the lower atmosphere or throughout the atmospheric column. Within the *Usoskin and Kovaltsov* [2012] range of uncertainty in spectral shape (indicated by the red arrow in Figure 16b), the Oct 89 x100, 29 Sep 89 x1000, and 20 Jan 05 x1000 events would take place at most, on average, once every 12,500 years (occurrence probability of $8 \times 10^{-5}$), implying that SPEs of this size are unlikely to have occurred during the Holocene.

## 6. Conclusion

This study sets out to quantify the fluence, duration, and energy spectra of solar proton events necessary to produce enhancements of $NO_y$ equivalent to nitrate spikes in the GISP2-H and BU ice cores that have been attributed to SPEs. We place these extreme events in the context of probability of occurrence distributions constructed using measured events, sunspot records, and cosmogenic radionuclide data. We conclude that events necessary to produce nitrate spikes in ice cores are beyond the probability of occurrence during the Holocene, confirming that nitrate ion spikes cannot be used as one-to-one, or even statistically representative, proxies for solar proton events.

Figure 17 summarizes our results. Figure 17a shows the total column $NO_y$ enhancement from large SPEs during the satellite era as well as high fluence hypothetical events, with the dashed horizontal line indicating a twofold increase. The top horizontal axis and shading show occurrence probabilities based on integral proton fluence greater than 30 MeV. An event 50 times the October 1989 series of events would be required to enhance $NO_y$ in the atmospheric column by a factor of two. It would take an event 100 times the October 1989 series of events to produce a threefold to fourfold increase in $NO_y$, similar in relative magnitude to the nitrate ion spikes in Greenland ice cores that some have attributed to SPEs. Probability distributions suggest that events of this size, with total integral fluence over 30 MeV of $\sim 2x10^{11}$ $cm^{-2}$ (Oct 89 x50) and $\sim 4x10^{11}$ $cm^{-2}$ (Oct 89 x100), are unlikely to have occurred during the Holocene ($\sim$10,000 years). Even rare millennial timescale events such as the event suggested by $^{14}C$ and $^{10}Be$ in 774-775 AD would be indiscernible in the nitrate record.

Figure 17b summarizes results from hard spectra (high energy) SPEs capable of producing odd nitrogen directly in the lower atmosphere. These graphs show $NO_y$ column enhancements from 0 to 30 km during the time period when PSCs are present, accounting for the SPE enhancement of odd nitrogen in the troposphere and lower stratosphere as well as the potential for enhanced nitrate deposition through uptake by gravitational settling of PSCs. Only the 20 Jan 05 x1000 and 20 Jan 05 10dx100 hypothetical SPEs produce a doubling of 0-30 km $NO_y$. Enhancements of $NO_x$ and $NO_y$ directly in the troposphere (0-10 km) are less than 1% for all simulations throughout the winter, while enhancements reaching the lower stratosphere (0-15 km) are less than 10% for $NO_x$ and less than 40% for $NO_y$ for the hypothetical SPEs. While events three orders of magnitude larger than the hard spectrum 20 Jan 2005 and 29 Sep 1989 SPEs show minimal significant increase in tropospheric $NO_x$, they are capable of producing a 130% and 60% increase of $NO_y$ from 0-30 km, respectively. Such hypothetical events, however, are well beyond the limits of the probability of occurrence during the Holocene.

These results show, in the framework of the present knowledge of atmospheric processes, that nitrate spikes in ice cores are not associated with individual SPEs. We conclude that nitrate ions in ice cores cannot, therefore, be used as proxies to study the frequency and strength of historical SPEs. Future efforts to analyze new cores for higher resolution nitrate spikes in the search for statistics involving SPEs are unwarranted. The results substantiate the argument by *Wolff et al.* [2012] that the 1859 Carrington SPE is not observed in ice cores and support the exclusion of nitrate as a proxy in the study of the frequency of extreme solar events by *Schrijver et al.* [2012]. Nitrate ion spikes observed in ice cores are not suitable proxies for SPEs.

## Acknowledgements

This work was supported by NSF grant 1135432 to the University of New Hampshire. We would like to acknowledge high-performance computing support from Yellowstone (ark:/85065/d7wd3xhc) provided by NCAR's Computational and Information Systems Laboratory, sponsored by the National Science Foundation. The CESM project is supported by the National Science Foundation and the Office of Science (BER) of the U.S. Department of Energy. We also acknowledge the support of NASA grant NNX14AH54G to the University of Colorado. We thank Colin Joyce for providing PREDDICS model results as well as Marco Calisto, Eugene Rozanov, and Ilya Usoskin and for providing ion pair production rates for the Carrington-like events. The model data used to produce the analysis and figures for this study are available upon request from the corresponding author. We thank the reviewers of the manuscript for their valuable comments and suggestions.

## References

Adriani, O. et al. (2011), PAMELA Measurements of Cosmic-ray Proton and Helium Spectra, *Science*, *332*(6025), 69–72, doi:10.1126/science.1199172.

Aguilar, M. et al. (2015), Precision Measurement of the Proton Flux in Primary Cosmic Rays from Rigidity 1 GV to 1.8 TV with the Alpha Magnetic Spectrometer on the International Space Station, *Phys. Rev. Lett.*, *114*(17), 171103, doi:10.1103/PhysRevLett.114.171103.

Atwell, W., A. Tylka, W. Dietrich, F Badavi, and K. Rojdev (2011), Spectral Analyses and Radiation Exposures from Several Ground-Level Enhancement (GLE) Solar Proton Events: A Comparison of Methodologies, 41st International Conference on Environmental Systems, 17-21 July 2011.

Aulanier, G., P. Démoulin, C. J. Schrijver, M. Janvier, E. Pariat, and B. Schmieder (2013), The standard flare model in three dimensions. II. Upper limit on solar flare energy, *Astron. Astrophys.*, *549*, A66, doi:10.1051/0004-6361/201220406.

Band, D. et al. (1993), BATSE observations of gamma-ray burst spectra. I - Spectral diversity, *Astrophys. J.*, *413*, 281–292, doi:10.1086/172995.

Barnard, L., M. Lockwood, M. A. Hapgood, M. J. Owens, C. J. Davis, and F. Steinhilber (2011), Predicting space climate change, *Geophys. Res. Lett.*, *38*(16), L16103, doi:10.1029/2011GL048489.

Bazilevskaya, G. A. et al. (2008), Cosmic Ray Induced Ion Production in the Atmosphere, *Space Sci Rev*, *137*(1-4), 149–173, doi:10.1007/s11214-008-9339-y.

Beer, J., K. McCracken, and R. von Steiger (2012), *Cosmogenic Radionuclides*, Physics of Earth and Space Environments, Springer Berlin Heidelberg, Berlin, Heidelberg.

Brakebusch, M., C. E. Randall, D. E. Kinnison, S. Tilmes, M. L. Santee, and G. L. Manney (2013), Evaluation of Whole Atmosphere Community Climate Model simulations of ozone during Arctic winter 2004–2005, *J. Geophys. Res. Atmos.*, *118*(6), 2673–2688, doi:10.1002/jgrd.50226.

Calisto, M., P. T. Verronen, E. Rozanov, and T. Peter (2012), Influence of a Carrington-like event on the atmospheric chemistry, temperature and dynamics, *Atmos. Chem. Phys.*, *12*(18), 8679–8686, doi:10.5194/acp-12-8679-2012.

Calisto, M., I. Usoskin, and E. Rozanov (2013), Influence of a Carrington-like event on the atmospheric chemistry, temperature and dynamics: revised, *Environ. Res. Lett.*, *8*(4), 045010, doi:10.1088/1748-9326/8/4/045010.

Cliver, E. W., A. J. Tylka, W. F. Dietrich, and A. G. Ling (2014), On a Solar Origin for the Cosmogenic Nuclide Event of 775 A.D., *Ap J*, *781*(1), 32, doi:10.1088/0004-637X/781/1/32.

Crosby, N. B. (2009), Solar extreme events 2005–2006: Effects on near-Earth space systems and interplanetary systems, *Adv. Space Res.*, *43*(4), 559–564, doi:10.1016/j.asr.2008.09.004.

Crutzen, P. J., I. S. A. Isaksen, and G. C. Reid (1975), Solar Proton Events: Stratospheric Sources of Nitric Oxide, *Science*, *189*(4201), 457–459, doi:10.1126/science.189.4201.457.

Dibb, J. E., and J.-L. Jaffrezo (1997), Air-snow exchange investigations at Summit, Greenland: An overview, *J. Geophys. Res.*, *102*(C12), 26795–26807, doi:10.1029/96JC02303.

Dibb, J. E., and S. I. Whitlow (1996), Recent climate anomalies and their impact on snow chemistry at South Pole, 1987-1994, *Geophys. Res. Lett.*, *23*(10), 1115–1118, doi:10.1029/96GL01039.

Dibb, J. E., E. Scheuer, M. Avery, J. Plant, and G. Sachse (2006), In situ evidence for renitrification in the Arctic lower stratosphere during the polar aura validation experiment (PAVE), *Geophys. Res. Lett.*, *33*(12), L12815, doi:10.1029/2006GL026243.

Duderstadt, K. A., J. E. Dibb, C. H. Jackman, C. E. Randall, S. C. Solomon, M. J. Mills, N. A. Schwadron, and H. E. Spence (2014), Nitrate deposition to surface snow at Summit, Greenland, following the 9 November 2000 solar proton event, *J. Geophys. Res. Atmos.*, *119*(11), 2013JD021389, doi:10.1002/2013JD021389.

Dunkerton, T. J., and D. P. Delisi (1986), Evolution of potential vorticity in the winter stratosphere of January-February 1979, *J. Geophys. Res.*, *91*(D1), 1199–1208, doi:10.1029/JD091iD01p01199.

Ellison, D. C., and R. Ramaty (1985), Shock acceleration of electrons and ions in solar flares, *Astrophy. J.*, *298*, 400–408, doi:10.1086/163623.

Emmons, L. K. et al. (2010), Description and evaluation of the Model for Ozone and Related chemical Tracers, version 4 (MOZART-4), *Geosci. Model Dev.*, *3*(1), 43–67, doi:10.5194/gmd-3-43-2010.

Feynman, J., T. P. Armstrong, L. Dao-Gibner, and S. Silverman (1990), New interplanetary proton fluence model, *J. Spacecraft Rockets*, *27*(4), 403–410, doi:10.2514/3.26157.

Fromm, M., J. Alfred, and M. Pitts (2003), A unified, long-term, high-latitude stratospheric aerosol and cloud database using SAM II, SAGE II, and POAM II/III data: Algorithm description, database definition, and climatology, *Journal of Geophysical Research (Atmospheres)*, *108*, 4366, doi:10.1029/2002JD002772.

Funke, B. et al. (2011), Composition changes after the "Halloween" solar proton event: the High Energy Particle Precipitation in the Atmosphere (HEPPA) model versus MIPAS data intercomparison study, *Atmos. Chem. Phys.*, *11*(17), 9089–9139, doi:10.5194/acp-11-9089-2011.

Garcia, R. R., D. R. Marsh, D. E. Kinnison, B. A. Boville, and F. Sassi (2007), Simulation of secular trends in the middle atmosphere, 1950–2003, *J. Geophys. Res.*, *112*(D9), D09301, doi:10.1029/2006JD007485.

Gopalswamy, N., H. Xie, S. Yashiro, S. Akiyama, P. Mäkelä, and I. G. Usoskin (2012), Properties of Ground Level Enhancement Events and the Associated Solar Eruptions during Solar Cycle 23, *arXiv:1205.0688 [astro-ph]*.

Jackman, C. H., J. E. Frederick, and R. S. Stolarski (1980), Production of odd nitrogen in the stratosphere and mesosphere: An intercomparison of source strengths, *J. Geophys. Res.*, *85*(C12), 7495–7505, doi:10.1029/JC085iC12p07495.

Jackman, C. H., A. R. Douglass, R. B. Rood, R. D. McPeters, and P. E. Meade (1990), Effect of solar proton events on the middle atmosphere during the past two solar cycles as computed using a two-dimensional model, *J. Geophys. Res.*, *95*(D6), 7417–7428, doi:10.1029/JD095iD06p07417.

Jackman, C. H., M. T. DeLand, G. J. Labow, E. L. Fleming, D. K. Weisenstein, M. K. W. Ko, M. Sinnhuber, and J. M. Russell (2005), Neutral atmospheric influences of the solar proton events in October–November 2003, *J. Geophys. Res.*, *110*(A9), A09S27, doi:10.1029/2004JA010888.

Jackman, C. H. et al. (2008), Short- and medium-term atmospheric constituent effects of very large solar proton events, *Atmos. Chem. Phys.*, *8*(3), 765–785, doi:10.5194/acp-8-765-2008.

Jackman, C. H., D. R. Marsh, F. M. Vitt, R. R. Garcia, C. E. Randall, E. L. Fleming, and S. M. Frith (2009), Long-term middle atmospheric influence of very large solar proton events, *J. Geophys. Res.*, *114*(D11), D11304, doi:10.1029/2008JD011415.

Jackman, C. H. et al. (2011), Northern Hemisphere atmospheric influence of the solar proton events and ground level enhancement in January 2005, *Atmos. Chem. Phys.*, *11*(13), 6153–6166, doi:10.5194/acp-11-6153-2011.

Jackman, C. H., C. E. Randall, V. L. Harvey, S. Wang, E. L. Fleming, M. Lopez-Puertas, B. Funke, and P. F. Bernath (2014), Middle Atmospheric Changes Caused by the January and March 2012 Solar Proton Events, *Atmos. Chem. Phys.*, *14*(2), 1025–1038, doi:10.5194/acp-14-1025-2014.

Joyce, C. J., Schwadron, N. A., Wilson, J. K., Spence, H. E., Kasper, J. C., Golightly, M., Blake, J. B., Mazur, J., Townsend, L. W., Case, A. W., Semones, E., Smith, S., and Zeitlin, C. J.,  Validation of PREDICCS using LRO/CRaTER observations during three major solar events in 2012,  *Space Weather*,  11,  350,  2013

Kepko, L., H. Spence, D. F. Smart, and M. A. Shea (2009), Interhemispheric observations of impulsive nitrate enhancements associated with the four large ground-level solar cosmic ray events (1940–1950), *J. Atmos. Sol-Terr. Phys.*, *71*(17–18), 1840–1845, doi:10.1016/j.jastp.2009.07.002.

Kim, M.-H. Y., M. J. Hayat, A. H. Feiveson, and F. A. Cucinotta (2009), Prediction of frequency and exposure level of solar particle events, *Health Phys*, *97*(1), 68–81, doi:10.1097/01.HP.0000346799.65001.9c.

Kinnison, D. E. et al. (2007), Sensitivity of chemical tracers to meteorological parameters in the MOZART-3 chemical transport model, *J. Geophys. Res.*, *112*(D20), D20302, doi:10.1029/2006JD007879.

Kovaltsov, G. A., and I. G. Usoskin (2014), Occurrence Probability of Large Solar Energetic Particle Events: Assessment from Data on Cosmogenic Radionuclides in Lunar Rocks, *Sol. Phys.*, *289*(1), 211–220, doi:10.1007/s11207-013-0333-5.

Kovaltsov, G. A., I. G. Usoskin, E. W. Cliver, W. F. Dietrich, and A. J. Tylka (2014), Fluence Ordering of Solar Energetic Proton Events Using Cosmogenic Radionuclide Data, *Sol. Phys.*, *289*(12), 4691–4700, doi:10.1007/s11207-014-0606-7.

Legrand, M. R., and S. Kirchner (1990), Origins and variations of nitrate in south polar precipitation, *J. Geophys. Res.*, *95*(D4), 3493–3507, doi:10.1029/JD095iD04p03493.

Legrand, M., and P. Mayewski (1997), Glaciochemistry of polar ice cores: A review, *Rev. Geophys.*, *35*(3), 219–243, doi:10.1029/96RG03527.

Legrand, M. R., F. Stordal, I. S. A. Isaksen, and B. Rognerud (1989), A model study of the stratospheric budget of odd nitrogen, including effects of solar cycle variations, *Tellus B*, *41B*(4), 413–426, doi:10.1111/j.1600-0889.1989.tb00318.x.

Li, G., G. P. Zank, and W. K. M. Rice (2003), Energetic particle acceleration and transport at coronal mass ejection–driven shocks, *J. Geophys. Res.*, *108*(A2), 1082, doi:10.1029/2002JA009666.

López-Puertas, M., B. Funke, S. Gil-López, T. von Clarmann, G. P. Stiller, M. Höpfner, S. Kellmann, H. Fischer, and C. H. Jackman (2005), Observation of NOx enhancement and ozone depletion in the Northern and Southern Hemispheres after the October–November 2003 solar proton events, *J. Geophys. Res.*, *110*(A9), A09S43, doi:10.1029/2005JA011050.

Manney, G. L., M. L., Santee, L., Froidevaux, K., Hoppel, N. J., Livesey, and J. W. Waters (2006), EOS MLS observations of ozone loss in the 2004–2005 Arctic winter, Geophys. Res. Lett., 33, L04802, doi:10.1029/2005GL024494.

Marsh, D. R., M. J. Mills, D. E. Kinnison, J.-F. Lamarque, N. Calvo, and L. M. Polvani (2013), Climate Change from 1850 to 2005 Simulated in CESM1(WACCM), *J. Climate*, *26*(19), 7372–7391, doi:10.1175/JCLI-D-12-00558.1.

Mazur, J. E., G. M. Mason, B. Klecker, and R. E. McGuire (1992), The energy spectra of solar flare hydrogen, helium, oxygen, and iron - Evidence for stochastic acceleration, *Astrophys. J.*, *401*, 398–410, doi:10.1086/172071.

McCracken, K. G., G. A. M. Dreschhoff, E. J. Zeller, D. F. Smart, and M. A. Shea (2001), Solar cosmic ray events for the period 1561-1994: 1. Identification in polar ice, 1561-1950, *J. Geophys. Res.*, *106*, 21585–21598, doi:10.1029/2000JA000237.

McCracken, K. G., H. Moraal, and M. A. Shea (2012), The high-energy impulsive ground-level enhancement, *ApJ*, *761*(2), 101, doi:10.1088/0004-637X/761/2/101.

McGuire, R. E., and T. T. von Rosenvinge (1984), The energy spectra of solar energetic particles, *Adv. Space Res.*, *4*(2–3), 117–125, doi:10.1016/0273-1177(84)90301-6.

Mertens, C. J., M. M. Meier, S. Brown, R. B. Norman, and X. Xu (2013), NAIRAS aircraft radiation model development, dose climatology, and initial validation, *Space Weather*, *11*(10), 603–635, doi:10.1002/swe.20100.

Mewaldt, R. A., C. M. S. Cohen, A. W. Labrador, R. A. Leske, G. M. Mason, M. I. Desai, M. D. Looper, J. E. Mazur, R. S. Selesnick, and D. K. Haggerty (2005), Proton, helium, and electron spectra during the large solar particle events of October–November 2003, *J. Geophys. Res.*, *110*(A9), A09S18, doi:10.1029/2005JA011038.

Mewaldt, R. A., M. D. Looper, C. M. S. Cohen, D. K. Haggerty, A. W. Labrador, R. A. Leske, G. M. Mason, J. E. Mazur, and T. T. von Rosenvinge (2012), Energy Spectra, Composition, and Other Properties of Ground-Level Events During Solar Cycle 23, *Space Sci Rev*, *171*(1-4), 97–120, doi:10.1007/s11214-012-9884-2.

Mironova, I. A., and I. G. Usoskin (2013), Possible effect of extreme solar energetic particle events of September–October 1989 on polar stratospheric aerosols: a case study, *Atmos. Chem. Phys.*, *13*(17), 8543–8550, doi:10.5194/acp-13-8543-2013.

Miroshnichenko, L. I., and R. A. Nymmik (2014), Extreme fluxes in solar energetic particle events: Methodological and physical limitations, *Radiation Measurements*, *61*, 6–15, doi:10.1016/j.radmeas.2013.11.010.

Miroshnichenko, L. I., E. V. Vashenyuk, and J. A. Pérez-Peraza (2013), Solar cosmic rays: 70 years of ground-based observations, *Geomagn. Aeron.*, *53*(5), 541–560, doi:10.1134/S0016793213050125.

Miyake, F., K. Nagaya, K. Masuda, and T. Nakamura (2012), A signature of cosmic-ray increase in ad 774-775 from tree rings in Japan, *Nature*, *486*(7402), 240–242, doi:10.1038/nature11123.

Miyake, F., A. Suzuki, K. Masuda, K. Horiuchi, H. Motoyama, H. Matsuzaki, Y. Motizuki, K. Takahashi, and Y. Nakai (2015), Cosmic ray event of A.D. 774–775 shown in quasi-annual 10Be data from the Antarctic Dome Fuji ice core, *Geophys. Res. Lett.*, *42*(1), 2014GL062218, doi:10.1002/2014GL062218.

Mottl, D., and R. Nymmik (2007), The issues of reliability of solar energetic proton flux databases and models, *Adv. Space Res.*, *39*(8), 1355–1361, doi:10.1016/j.asr.2007.01.055.

Mottl, D. A., R. A. Nymmik, A. I. Sladkova (2001), Energy spectra of high-energy SEP event protons derived from statistical analysis of experimental data on a large set of events, in: Mohamed S. (Ed). American Institute of Physics, Conference Proceedings, vol. 552 (9). Spinger-Verlag, New York, pp. 1191–1196, doi:10.1063/1.1358071.

NASA (1996), GOES I-M DataBook DRL 101-08,

NASA (2006), GOES N Data Book CDRL PM-1-1-03,

Neale, R. B., J. Richter, S. Park, P. H. Lauritzen, S. J. Vavrus, P. J. Rasch, and M. Zhang (2013), The Mean Climate of the Community Atmosphere Model (CAM4) in Forced SST and Fully Coupled Experiments, *J. Climate*, *26*(14), 5150–5168, doi:10.1175/JCLI-D-12-00236.1.

Porter, H. S., C. H. Jackman, and A. E. S. Green (1976), Efficiencies for production of atomic nitrogen and oxygen by relativistic proton impact in air, *J. Chem. Phys.*, *65*(1), 154–167, doi:10.1063/1.432812.

Randall, C. E. et al. (2005), Stratospheric effects of energetic particle precipitation in 2003–2004, *Geophys. Res. Lett.*, *32*(5), L05802, doi:10.1029/2004GL022003.

Randall, C. E., V. L. Harvey, D. E. Siskind, J. France, P. F. Bernath, C. D. Boone, and K. A. Walker (2009), $NO_x$ descent in the Arctic middle atmosphere in early 2009, *Geophys. Res. Lett.*, *36*(18), L18811, doi:10.1029/2009GL039706.

Reedy, R. C. (2006), Solar-Proton Event-Integrated Fluences During the Current Solar Cycle, 37th Annual Lunar and Planetary Science Conference, 13-17 March 2006.

Rienecker, M. M. et al. (2011), MERRA: NASA's Modern-Era Retrospective Analysis for Research and Applications, *J. Climate*, *24*(14), 3624–3648, doi:10.1175/JCLI-D-11-00015.1.

Riley, P. (2012), On the probability of occurrence of extreme space weather events, *Space Weather*, *10*(2), S02012, doi:10.1029/2011SW000734.

Rodger, C. J., P. T. Verronen, M. A. Clilverd, A. Seppälä, and E. Turunen (2008), Atmospheric impact of the Carrington event solar protons, *J. Geophys. Res.*, *113*(D23), D23302, doi:10.1029/2008JD010702.

Röthlisberger, R. et al. (2002), Nitrate in Greenland and Antarctic ice cores: a detailed description of post-depositional processes, *Ann. Glaciol., 35*, 209–216, doi:10.3189/172756402781817220.

Rusch, D. W., J.-C. Gérard, S. Solomon, P. J. Crutzen, and G. C. Reid (1981), The effect of particle precipitation events on the neutral and ion chemistry of the middle atmosphere—I. Odd nitrogen, *Planet. Space Sci.*, *29*(7), 767–774, doi:10.1016/0032-0633(81)90048-9.

Schrijver, C. J. et al. (2012), Estimating the frequency of extremely energetic solar events, based on solar, stellar, lunar, and terrestrial records, *J. Geophys. Res.*, *117*(A8), A08103, doi:10.1029/2012JA017706.

Schwadron, N. (2012), Near-Real-Time Situational Awareness of Space Radiation Hazards, *Space Weather*, *10*(10), S10005, doi:10.1029/2012SW000860.

Shea, M. A., and D. F. Smart (1990), A summary of major solar proton events, *Sol Phys*, *127*(2), 297–320, doi:10.1007/BF00152170.

Shea, M. A., and D. F. Smart (2012), Space Weather and the Ground-Level Solar Proton Events of the 23rd Solar Cycle, *Space Sci Rev*, *171*(1-4), 161–188, doi:10.1007/s11214-012-9923-z.

Shea, M. A., D. F. Smart, K. G. McCracken, G. A. M. Dreschhoff, and H. E. Spence (2006), Solar proton events for 450 years: The Carrington event in perspective, *Adv. Space Res.*, *38*(2), 232–238, doi:10.1016/j.asr.2005.02.100.

Smart, D. F., and M. A. Shea (1999), Comment on the use of GOES solar proton data and spectra in solar proton dose calculations, *Radiation Measurements*, *30*(3), 327–335, doi:10.1016/S1350-4487(99)00059-1.

Smart, D. F., M. A. Shea, and K. G. McCracken (2006a), The Carrington event: Possible solar proton intensity–time profile, *Adv. Space Res.*, *38*(2), 215–225, doi:10.1016/j.asr.2005.04.116.

Smart, D. F., M. A. Shea, H. E. Spence, and L. Kepko (2006b), Two groups of extremely large >30 MeV solar proton fluence events, *Adv. Space Res.*, *37*(9), 1734–1740, doi:10.1016/j.asr.2005.09.008.

Smart, D. F., M. A. Shea, A. L. Melott, and C. M. Laird (2014), Low time resolution analysis of polar ice cores cannot detect impulsive nitrate events, *J. Geophys. Res.: Space Physics*, *119*(12), 9430–9440, doi:10.1002/2014JA020378.

Solomon, S., D. W. Rusch, J. C. Gérard, G. C. Reid, and P. J. Crutzen (1981), The effect of particle precipitation events on the neutral and ion chemistry of the middle atmosphere: II. Odd hydrogen, *Planet. Space Sci.*, *29*(8), 885–893, doi:10.1016/0032-0633(81)90078-7.

Sternheimer, R. M. (1959), Range-Energy Relations for Protons in Be, C, Al, Cu, Pb, and Air, *Phys. Rev.*, *115*(1), 137–142, doi:10.1103/PhysRev.115.137.

Thomas, B. C., A. L. Melott, K. R. Arkenberg, and B. R. Snyder II (2013), Terrestrial effects of possible astrophysical sources of an AD 774-775 increase in 14C production, *Geophys. Res. Lett.*, *40*(6), 1237–1240, doi:10.1002/grl.50222.

Townsend, L. W., E. N. Zapp, J., D.L. Stephens, and J. L. Hoff (2003), Carrington flare of 1859 as a prototypical worst-case solar energetic particle event, *IEEE T. Nucl. Sci.*, *50*(6), 2307–2309, doi:10.1109/TNS.2003.821602.

Townsend, L. W., D. L. Stephens Jr., J. L. Hoff, E. N. Zapp, H. M. Moussa, T. M. Miller, C. E. Campbell, and T. F. Nichols (2006), The Carrington event: Possible doses to crews in space from a comparable event, *Adv. Space Res.*, *38*(2), 226–231, doi:10.1016/j.asr.2005.01.111.

Tylka, A. J., and W. F. Dietrich (2009), A new and comprehensive analysis of proton spectra in ground-level enhanced (GLE) solar particle events, Proceedings of the 31st ICRC, Lódź.

Tylka, A. J., C. M. S. Cohen, W. F. Dietrich, M. A. Lee, C. G. Maclennan, R. A. Mewaldt, C. K. Ng, and D. V. Reames (2006), A Comparative Study of Ion

Characteristics in the Large Gradual Solar Energetic Particle Events of 2002 April 21 and 2002 August 24, *ApJS*, *164*(2), 536, doi:10.1086/503203.

Usoskin, I. G., and G. A. Kovaltsov (2006), Cosmic ray induced ionization in the atmosphere: Full modeling and practical applications, *J. Geophys. Res.*, *111*(D21), D21206, doi:10.1029/2006JD007150.

Usoskin, I. G., and G. A. Kovaltsov (2012), Occurrence of Extreme Solar Particle Events: Assessment from Historical Proxy Data, *ApJ*, *757*(1), 92, doi:10.1088/0004-637X/757/1/92.

Usoskin, I. G., O. G. Gladysheva, and G. A. Kovaltsov (2004), Cosmic ray-induced ionization in the atmosphere: spatial and temporal changes, *J. Atmos. Sol-Terr. Phys.*, *66*(18), 1791–1796, doi:10.1016/j.jastp.2004.07.037.

Usoskin, I. G., A. J. Tylka, G. A. Kovaltsov, and W. F. Dietrich (2009), Ionization effect of strong solar particle events: Low-middle atmosphere, *Proceedings of the 31st ICRC, Lodz, icrc0162*.

Usoskin, I. G., G. A. Kovaltsov, and I. A. Mironova (2010), Cosmic ray induced ionization model CRAC:CRII: An extension to the upper atmosphere, *J. Geophys. Res.*, *115*(D10), D10302, doi:10.1029/2009JD013142.

Usoskin, I. G., G. A. Kovaltsov, I. A. Mironova, A. J. Tylka, and W. F. Dietrich (2011), Ionization effect of solar particle GLE events in low and\ middle atmosphere, *Atmos. Chem. Phys.*, *11*(5), 1979–1988, doi:10.5194/acp-11-1979-2011.

Usoskin, I. G., B. Kromer, F. Ludlow, J. Beer, M. Friedrich, G. A. Kovaltsov, S. K. Solanki, and L. Wacker (2013), The AD775 cosmic event revisited: the Sun is to blame, *Astron. Astrophys.*, *552*, L3, doi:10.1051/0004-6361/201321080.

Vashenyuk, E. V., Y. V. Balabin, and B. B. Gvozdevsky (2011), Features of relativistic solar proton spectra derived from ground level enhancement events (GLE) modeling, *Astrophys. Space Sci. Trans.*, *7*(4), 459–463, doi:10.5194/astra-7-459-2011.

Vitt, F. M., T. P. Armstrong, T. E. Cravens, G. A. M. Dreschhoff, C. H. Jackman, and C. M. Laird (2000), Computed contributions to odd nitrogen concentrations in the Earth's polar middle atmosphere by energetic charged particles, *J. Atmos. Sol-Terr. Phys.*, *62*(8), 669–683, doi:10.1016/S1364-6826(00)00048-1.

Wang, R. (2009), Did the 2000 July 14 solar flare accelerate protons to $\geqslant$40 GeV?, *Astropart. Phys.*, *31*(2), 149–155, doi:10.1016/j.astropartphys.2008.12.005.

Wang, R., and J. Wang (2006), Spectra and solar energetic protons over 20 GeV in Bastille Day event, *Astropart. Phys.*, *25*(1), 41–46, doi:10.1016/j.astropartphys.2005.11.002.

Webber, W. R., P. R. Higbie, and K. G. McCracken (2007), Production of the cosmogenic isotopes 3H, 7Be, 10Be, and 36Cl in the Earth's atmosphere by solar and galactic cosmic rays, *J. Geophys. Res.*, *112*(A10), A10106, doi:10.1029/2007JA012499.

Wolff, E. W., A. E. Jones, S. J.-B. Bauguitte, and R. A. Salmon (2008), The interpretation of spikes and trends in concentration of nitrate in polar ice cores, based on evidence from snow and atmospheric measurements, *Atmos. Chem. Phys.*, *8*(18), 5627–5634, doi:10.5194/acp-8-5627-2008.

Wolff, E. W., M. Bigler, M. a. J. Curran, J. E. Dibb, M. M. Frey, M. Legrand, and J. R. McConnell (2012), The Carrington event not observed in most ice core nitrate records, *Geophys. Res. Lett.*, *39*(8), L08503, doi:10.1029/2012GL051603.

Xapsos, M. A., and J. L. B. (2001), Characterizing solar proton energy spectra for radiation effects applications, *IEEE T. Nucl. Sci.* (6), 2218 – 2223, doi:10.1109/23.903756.

Zeller, E. J., and G. A. M. Dreschhoff (1995), Anomalous nitrate concentrations in polar ice cores—Do they result from solar particle injections into the polar atmosphere?, *Geophys. Res. Lett.*, *22*(18), 2521–2524, doi:10.1029/95GL02560.

Table 1. Comparison of Solar Proton Events from 1956 to 2012

| Solar Proton Event Date | $NO_y$ Prod[1] Gmol | GLE Number[2] | GLE Peak[3] % | $F_{30}$[4] $cm^{-2}$ | $F_{200}$[5] $cm^{-2}$ |
|---|---|---|---|---|---|
| 23 Feb 1956 | -- | 5 | 5000 | 1.8E+09 | 1.2E+08 |
| 10-17 Jul 1959 | -- | 7 | 18 | 2.3E+09 | 1.6E+07 |
| 12-20 Nov 1960 | -- | 10,11,12 | 150,100,-- | 9.0E+09 | 1.0E+08 |
| 2-7 Sep 1966 | 2.0 | -- | -- | -- | -- |
| 28 Jan-1 Feb 1967 | 1.6 | 16 | 30 | -- | 4.3E+06 |
| 2-10 Aug 1972 | 6.0 | 24,25 | 79,17 | 5.0E+09 | 1.7E+07 |
| 13-26 Aug 1989 | 3.0 | 41 | 24 | 1.5E+09 | 5.1E+06 |
| 29 Sep-3 Oct 1989 | 1.7 | 42 | 395 | 1.8E+08 | 3.1E+07 |
| 19-27 Oct 1989 | 11 | 43,44,45 | 90,190,95 | 4.3E+09 | 9.3E+07 |
| 21-28 May 1990 | -- | 47,48,49,50 | 24,50,--,-- | -- | 6.4E+06 |
| 11-15 Jun 1991 | -- | 51,52 | 12,57 | -- | 8.3E+06 |
| 14-16 Jul 2000 | 5.8 | 59 | 80 | 4.3E+09 | 3.4E+07 |
| 9-11 Nov 2000 | 4.3 | -- | -- | 3.1E+09 | -- |
| 15-18 April 2001 | -- | 60,61 | 230,26 | 1.5E+08 | 9.3E+06 |
| 24-30 Sep 2001 | 3.3 | -- | -- | 1.2E+09 | -- |
| 5-7 Nov 2001 | 5.3 | 62 | -- | 3.4E+09 | 1.2E+07 |
| 23-25 Nov 2001 | 2.8 | -- | -- | 8.5E+08 | -- |
| 28-31 Oct 2003 | 5.6 | 65,66 | 45,35 | 3.3E+09 | 2.3E+07 |
| 15-23 Jan 2005 | 1.8 | 68,69 | --,5500 | 1.0E+09 | 2.4E+07 |
| 23-30 Jan 2012 | 1.9 | -- | -- | 5.5E+08 | -- |
| 7-11 Mar 2012 | 2.1 | -- | -- | 9.6E+08 | -- |

1. Computed $NO_y$ production in middle atmosphere (Gigamoles). *Jackman et al.* [2008, 2009, 2014].
2. GLE number is associated with the number assigned to each GLE recorded by neutron monitors since 28 February 1942.
3. Maximum peak percent increase for corresponding GLEs from Neutron Monitor Database as describe by *McCracken et al.* [2012]. The early data are mostly five-minute averages, with one-minute averages more frequent after 1990. Values less than 10% are represented by "--".
4. Fluence >30 MeV. *Shea et al.* [2006]; *Smart et al.,* [2006b]; *Reedy et al.* [2006]; *Webber et al.* [2007], *Reedy* [2015]. Values less than 1 x $10^8$ $cm^{-2}$ or unknown to the authors are represented by "--".
5. Fluence >200 MeV. *Kovaltsov et al.* [2014]. Values less than 1 x $10^6$ $cm^{-2}$ or unknown to the authors are represented by "--".

Table 2. Simulation Runs

| Observed | | | Hypothetical High Fluence | Hypothetical Hard Spectra | Carrington-like |
|---|---|---|---|---|---|
| 29 | Sep | 1989 | 19-27 Oct 1989 x10 | 29 Sep 1989 x100 | 23 Feb 1956 (scaled to $1.9 \times 10^{10}$ cm$^{-2}$ $F_{30}$) |
| 19-27 | Oct | 1989 | 19-27 Oct 1989 x50 | 29 Sep 1989 x1000 | |
| 14 | Jul | 2000 | 19-27 Oct 1989 x100 | 29 Sep 1989 10dx100 | |
| 9 | Nov | 2000 | 19-27 Oct 1989 10dx10 | 20 Jan 2005 x100 | 4 Aug 1972 (scaled to $1.9 \times 10^{10}$ cm$^{-2}$ $F_{30}$) |
| 26-31 | Oct | 2003 | | 20 Jan 2005 x1000 | |
| 15-23 | Jan | 2005 | | 20 Jan 2005 10dx100 | |
| 23 | Jul | 2012 | | | |

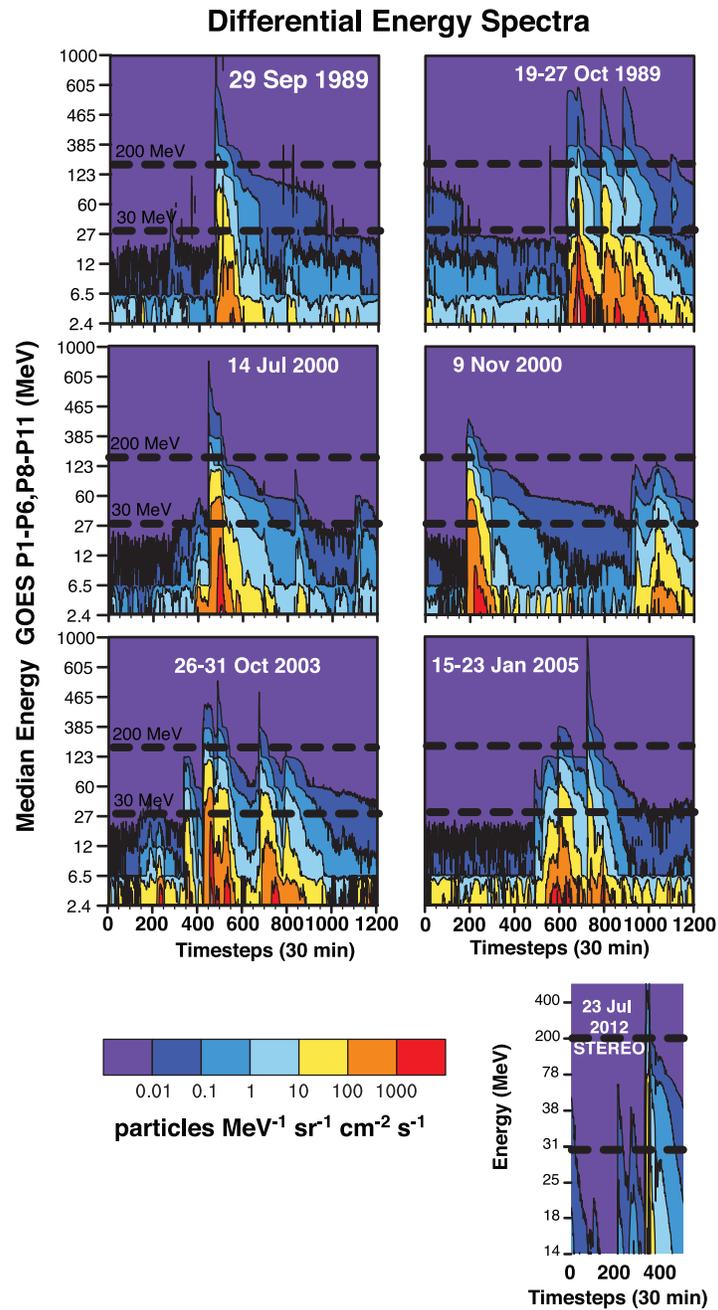

Figure 1. Differential energy flux as a function of GOES and STEREO energy level and time. Note that vertical axis refers to energy channel and is neither linear nor logarithmic. Dashed lines denote 30 MeV and 200 MeV energy levels corresponding to Table 1.

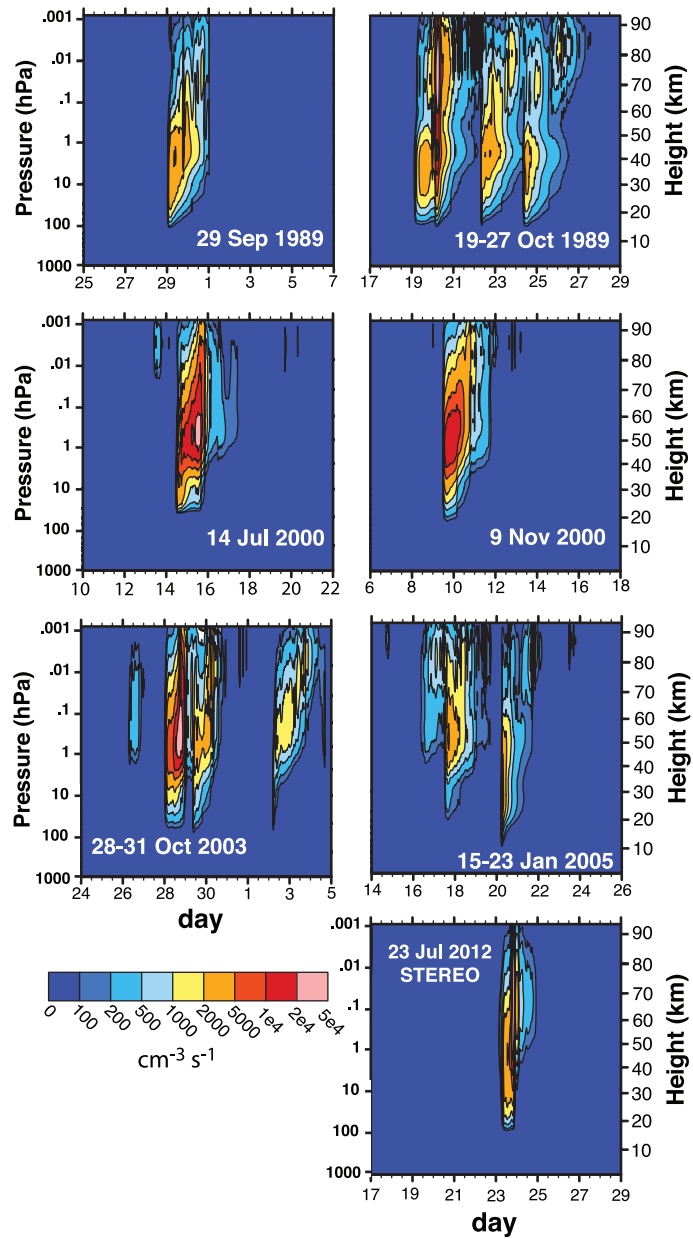

Figure 2. Ion pair production rates (cm$^{-3}$ s$^{-1}$) as a function of altitude (pressure and height) and time (day of the month).

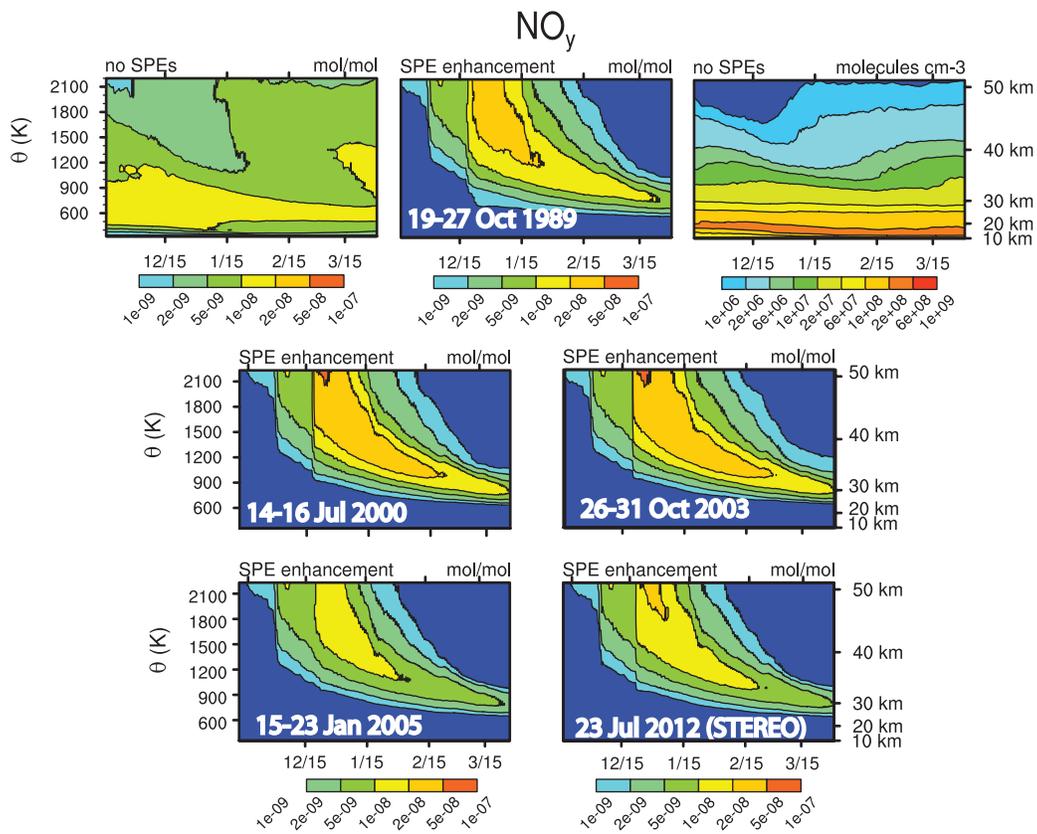

Figure 3. Time evolution of the Arctic vortex-averaged enhancement of $NO_y$ (mole ratios) for SPEs placed in the winter of 2004-2005. Top left: Background $NO_y$ (mole ratios) with no SPEs. Top right: Background $NO_y$ (number density) with no SPEs. Dates are given as month/day.

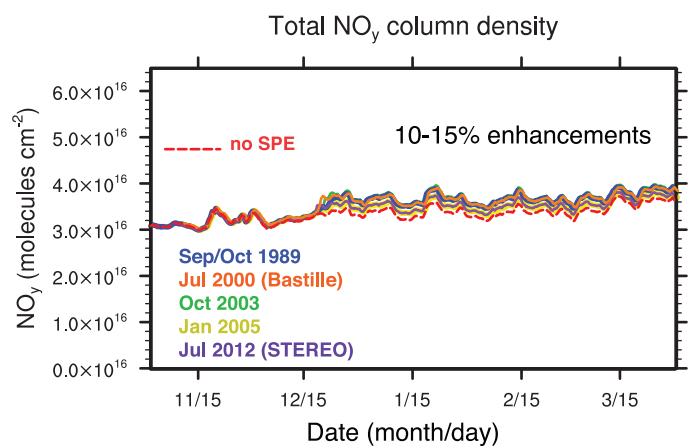

Figure 4. Arctic vortex-averaged total $NO_y$ column density (molecules cm$^{-2}$) for several SPEs placed in the 2004-2005 winter. All events begin on 18 December 2004.

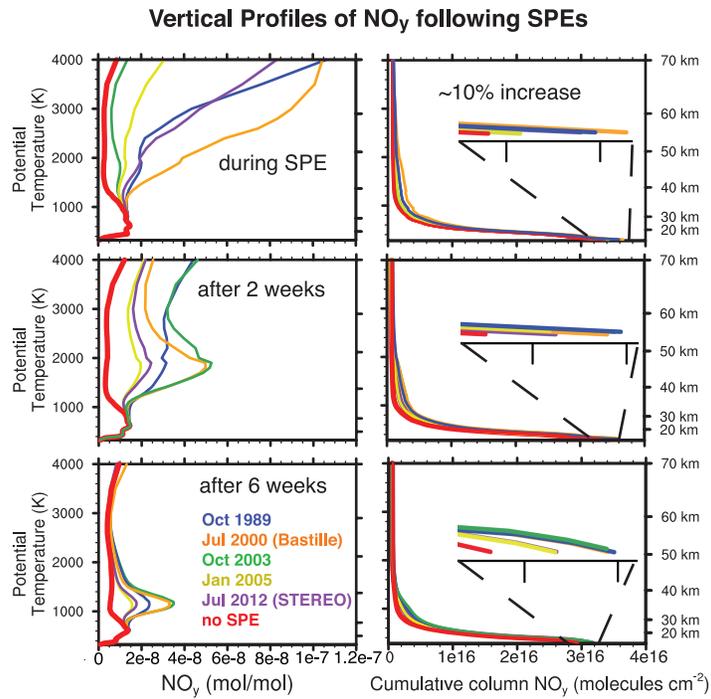

Figure 5. Profiles of Arctic vortex-averaged $NO_y$ (mol/mol) and cumulative column $NO_y$ (molecules $cm^{-2}$) for several high fluence SPEs placed in the winter of 2004-2005. Top: during the SPEs. Middle: two weeks following SPEs. Bottom: six weeks following SPEs.

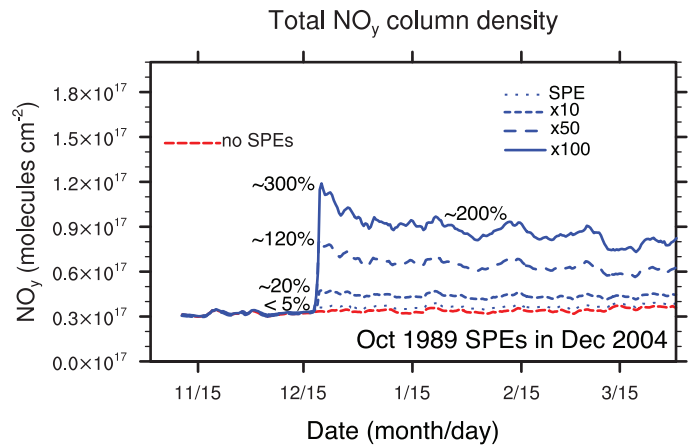

Figure 6. Arctic vortex-averaged total $NO_y$ column density (molecules cm$^{-2}$) throughout the 2004-2005 winter, comparing simulations without SPEs (red) and scaled simulations with SPEs (blue) based on the Oct 1989 events.

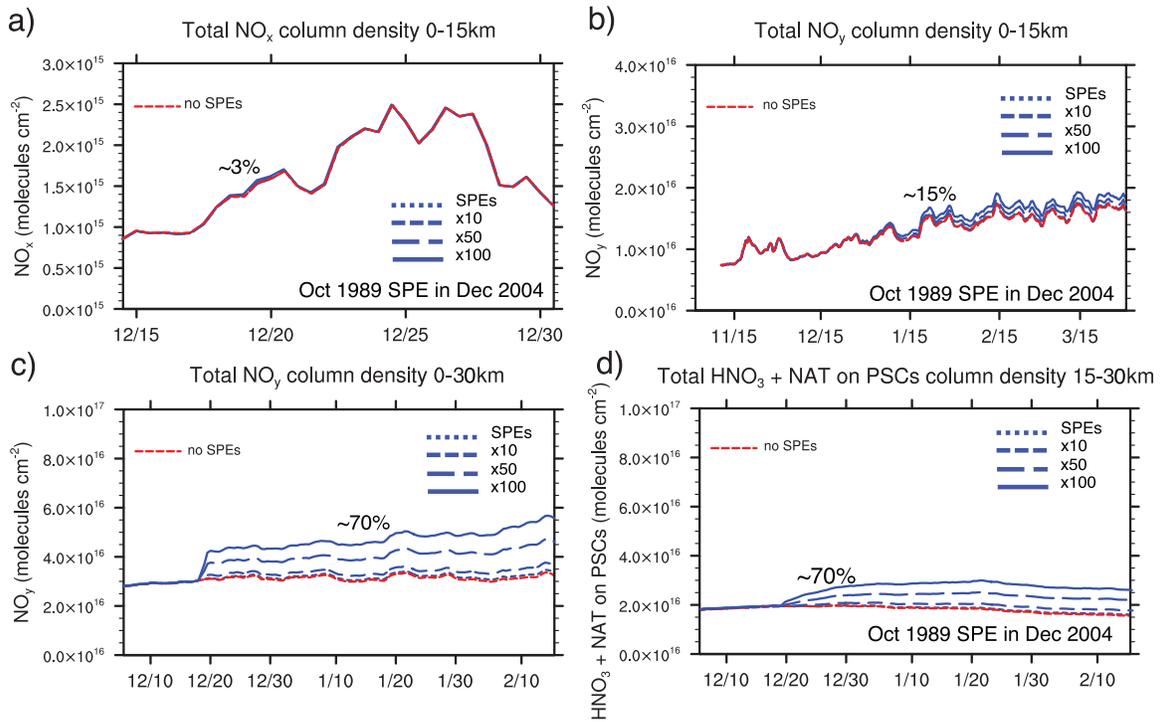

Figure 7. Arctic vortex-averaged column density (molecules cm$^{-2}$) of: a) NO$_x$ from 0-15 km, b) NO$_y$ from 0-15 km, c) NO$_y$ from 0-30 km, and d) HNO$_3$ from 15-30 km. Hypothetical events based on the Oct 1989 SPEs.

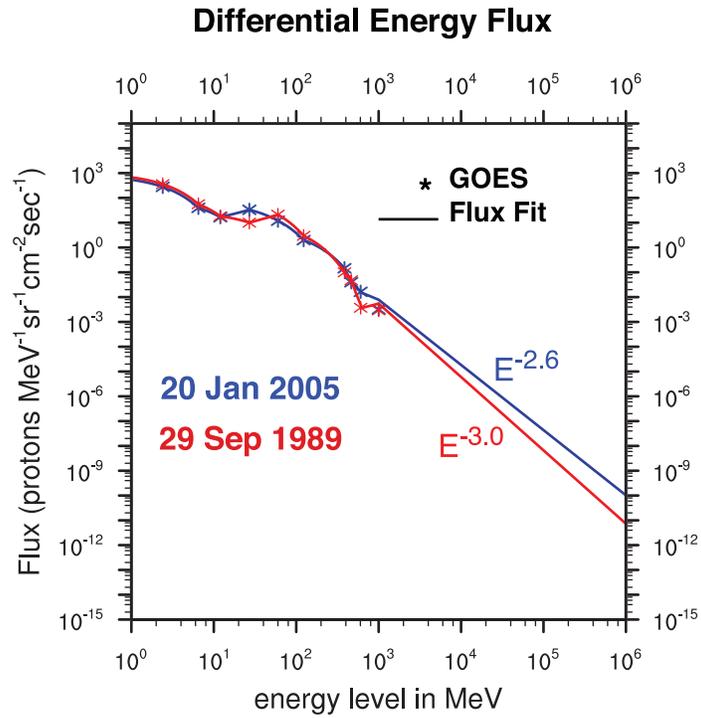

Figure 8. Spectral fits for 20 Jan 2005 and 29 Sep 1989 SPEs based on GOES EPS and HEPAD observations. Spectral indices are -3.0 for the 29 Sep 1989 SPE and -2.6 for the 20 Jan 2005 SPE.

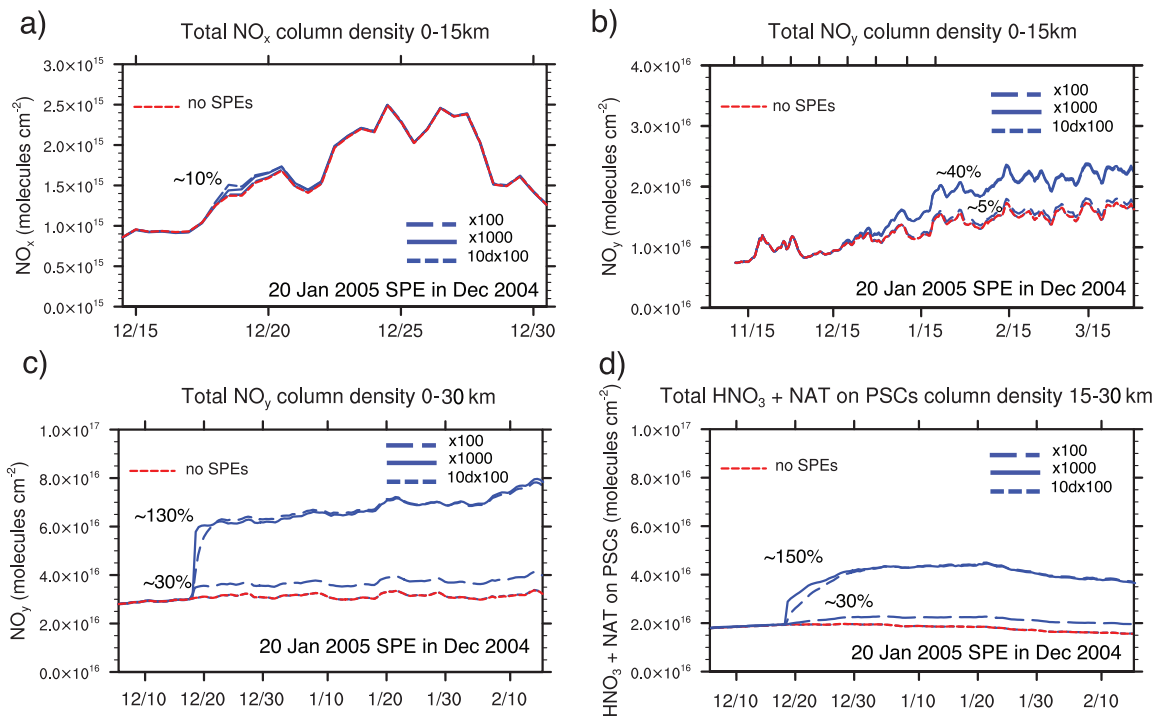

Figure 9. Arctic vortex-averaged column density (molecules cm$^{-2}$) of: a) NO$_x$ from 0-15 km, b) NO$_y$ from 0-15 km, c) NO$_y$ from 0-30 km, and d) HNO$_3$ from 15-30 km. Hypothetical events based on the 20 Jan 2005 SPEs.

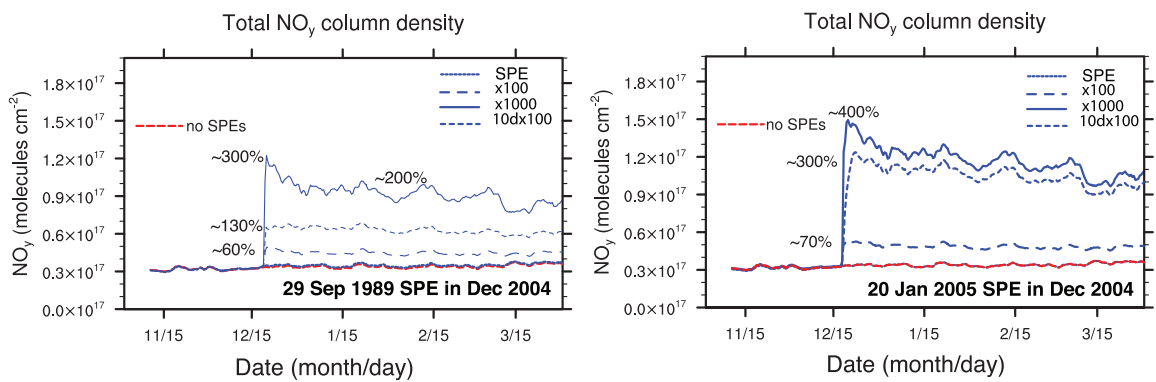

Figure 10. Arctic vortex-averaged total NO$_y$ column density (molecules cm$^{-2}$) throughout the 2004-2005 winter, comparing simulations without SPEs (red) and simulations with SPEs (blue) scaled from the 29 Sep 1989 and 20 Jan 2005 SPEs.

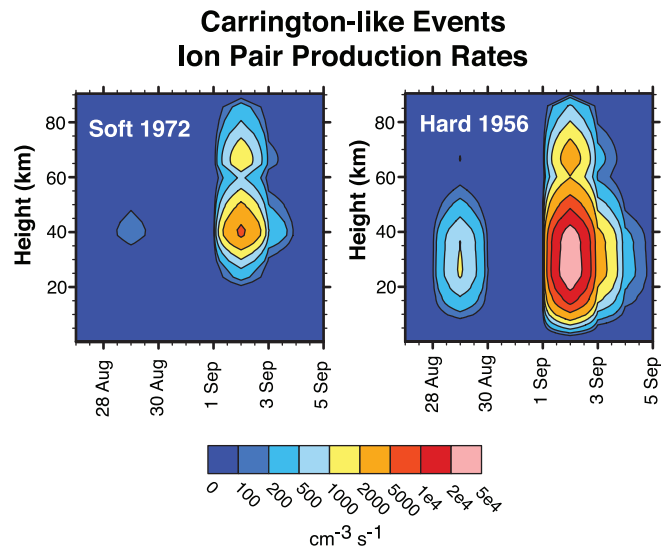

Figure 11. Ionization profiles for Carrington-like events as a function of time. Left: "Soft" Carrington-like event modeled after August 1972 SPE. Right: "Hard" Carrington-like event modeled after February 1956 SPE.

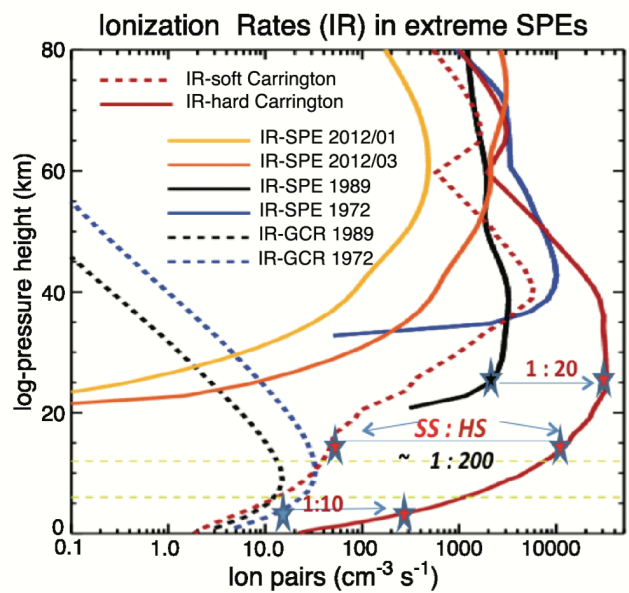

Figure 12. Comparison of ionization profiles of the "IR-soft" (Aug 1972-based) and "IR-hard" (Feb 1956-based) Carrington-like SPEs used in WACCM simulations with SPE profiles from Aug 1972, Oct 1989, Jan 2012, and Mar 2012 and GCR rates (annual mean) from 1972 and 1989. [*Usoskin et al.,* 2010, 2012; *Mertens et al.,* 2013; *Jackman et al.,* 2014]

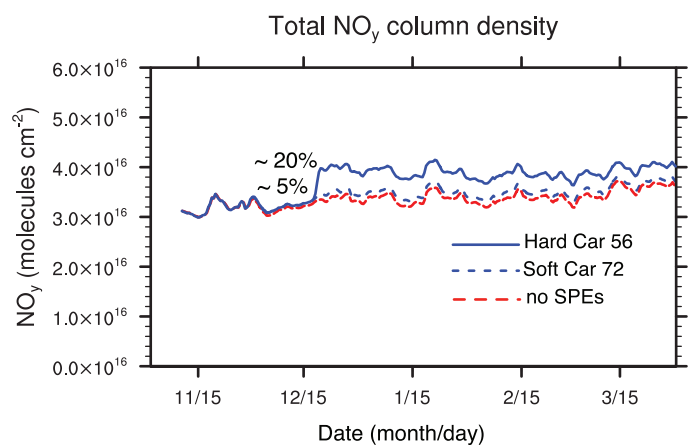

Figure 13. Arctic vortex-averaged total $NO_y$ column density (molecules $cm^{-2}$) throughout the 2004-2005 winter, comparing simulations without SPEs (red) and Carrington-like simulations based on the Aug 1972 SPE (Soft Car 72) and Feb 1956 SPE (Hard Car 56).

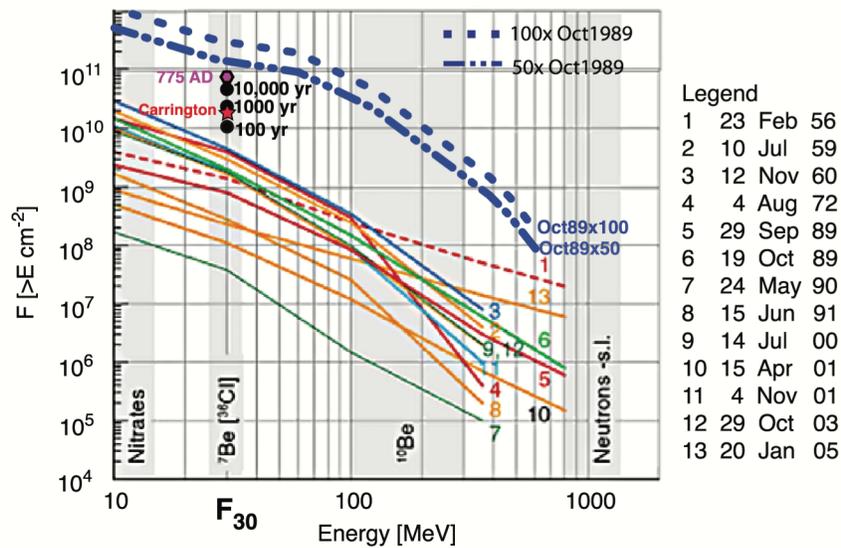

Figure 14. Total integral fluence spectra for SPEs from 1956-2005 adapted from *Webber et al.* [2007] and *Beer et al.* [2012]. Shaded regions indicate peak response energies for nitrates, [7]Be, [10]Be, and neutron monitors. Hypothetical events for 50x and 100x the October 1989 SPEs used in WACCM simulations are overlaid on the plot. Black circles indicate probability of occurrence from *Usoskin and Kovaltsov* [2012]. Red star identifies the estimated Carrington event fluence from *McCracken et al.* [2001] based on GISP2-H nitrate. Purple hexagon shows fluence for potential SPE in 774-775 [*Cliver et al.,* 2014].

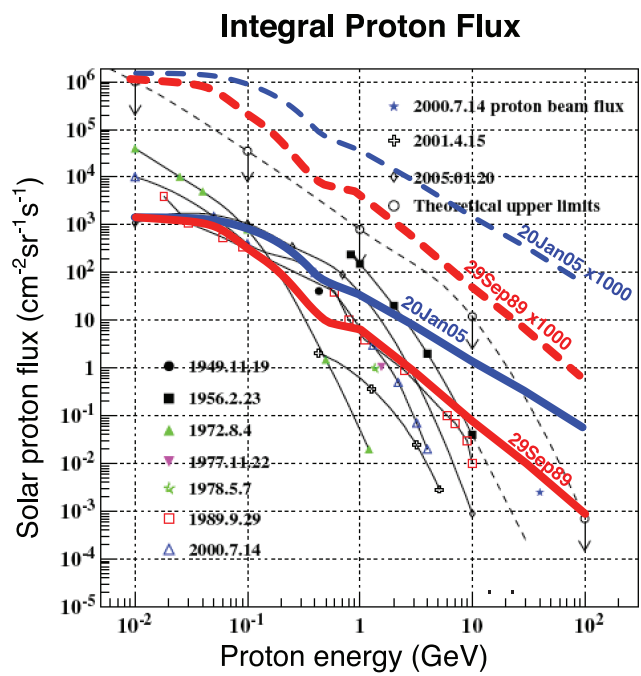

Figure 15. Integral flux of solar proton events during peak flux. Adapted from *Wang et al.* [2009]. Hard spectra events used in WACCM are overlaid in red (29Sep89 and 29Sep89x1000) and blue (20Jan05 and 20Jan05x1000).

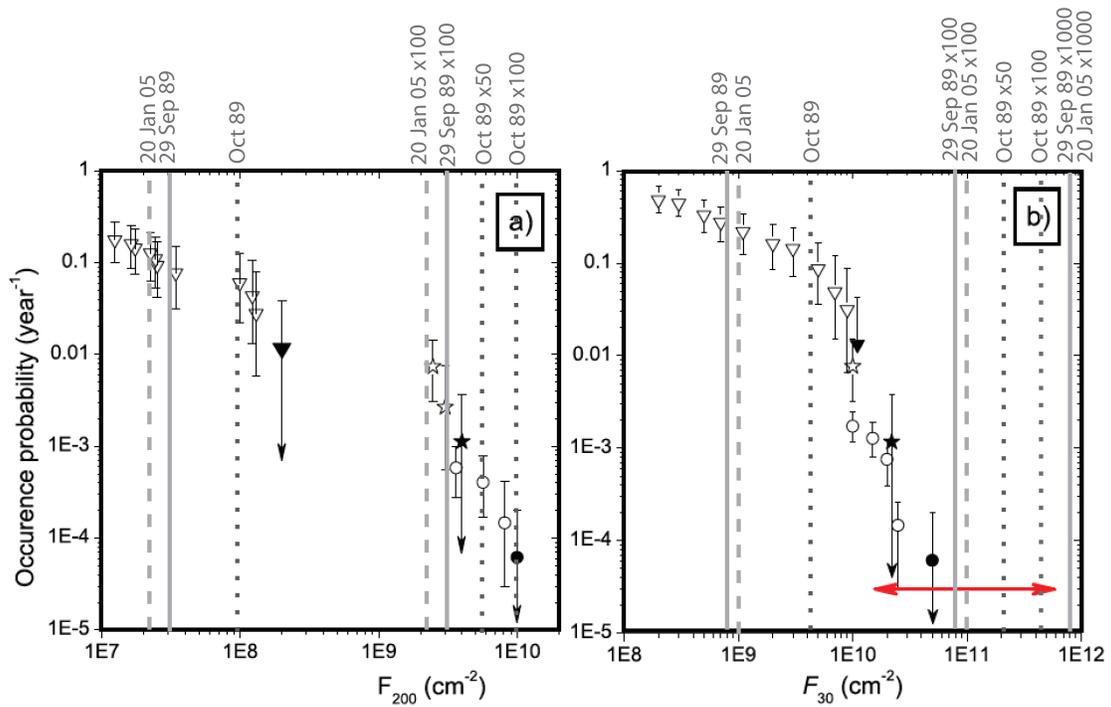

Figure 16. Probability of occurrence for SPEs having fluence integrated above a) 200 MeV ($F_{200}$) and b) 30 MeV ($F_{30}$). Adapted from *Usoskin and Kovaltsov* [2012] and *Kovaltsov et al.* [2014]. The red arrow indicates the range of uncertainty due to unknown spectral shape as in *Kovaltsov et al.* [2014]. Vertical gray lines represent the hypothetically extreme SPEs used in our WACCM simulations.

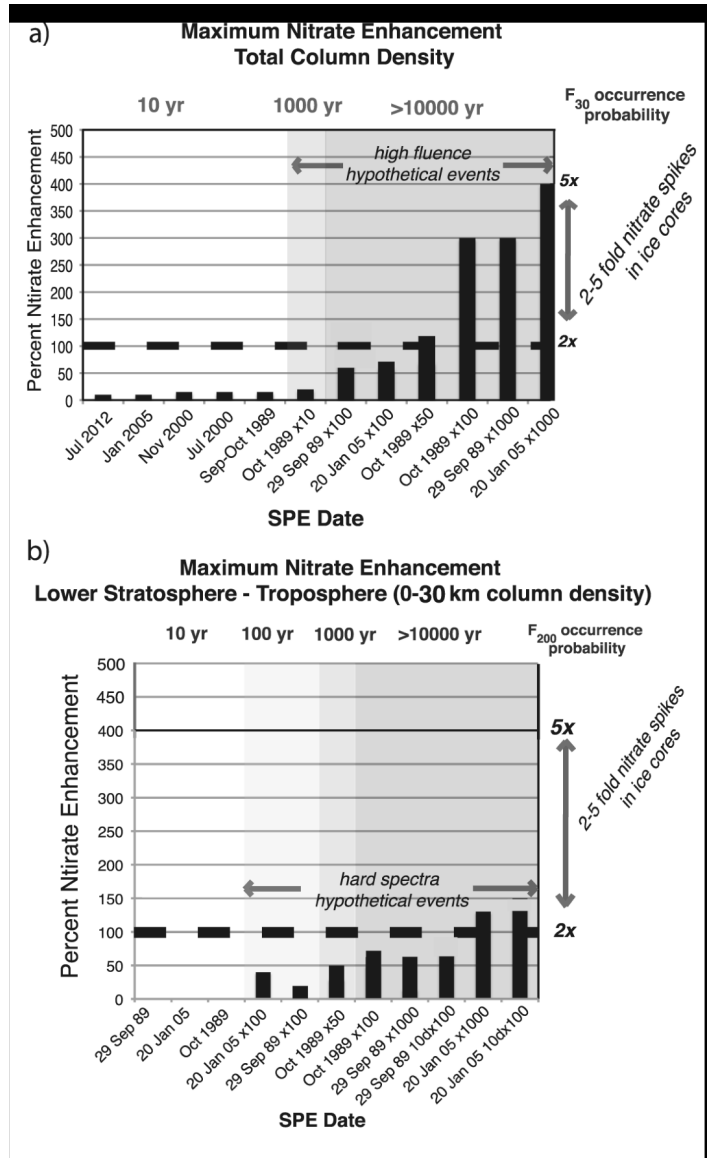

Figure 17. Nitrate enhancement resulting from SPEs measured by satellite and hypothetical events. Probability of occurrence from *Kovaltsov et al.* [2014] indicated on top horizontal axis and as shaded regions. a) Peak total column density of $NO_y$ for high fluence events with $F_{30}$ probability of occurrence. b) Peak 0-30 km $NO_y$ column density for hard spectrum events with $F_{200}$ probability of occurrence.